\documentclass[12pt]{article}
\pdfoutput=1
\usepackage{amsmath,amssymb,amsfonts,graphicx}
\usepackage{slashed,tabularx,cite}
\usepackage{mathrsfs}
\usepackage[all]{xy}

\usepackage[usenames,dvipsnames]{xcolor}
\usepackage[colorlinks=true, pdfstartview=FitV, linkcolor=black, 
	citecolor=black, urlcolor=black]{hyperref}

\newcommand{\leigh}{Robert G. Leigh}
\newcommand{\onkar}{Onkar Parrikar}
\newcommand{\kewang}{Kewang Jin}

\newcommand{\addressuiuc}{
	Department of Physics, University of Illinois, 1110 West Green St., Urbana IL 61801, U.S.A.
	}
\newcommand{\addresskitp}{
Kavli Institute for Theoretical Physics, University of California, Santa Barbara CA 93106, U.S.A.}
\newcommand{\thistitle}{
	Higher Spin Fronsdal Equations\\ from the Exact Renormalization Group
	}

\edef\marginnotetextwidth{\the\textwidth}

\newcommand\cc[1]{#1^{^{\kern-6pt \circ}}\kern2pt}

\newcommand{\pa}{\partial}

\newcommand{\beq}{\begin{equation}}
\newcommand{\eeq}{\end{equation}}
\newcommand{\beqn}{\begin{eqnarray}}
\newcommand{\eeqn}{\end{eqnarray}}

\def\dalemb#1#2{{\vbox{\hrule height .#2pt
\hbox{\vrule width.#2pt height#1pt \kern#1pt
\vrule width.#2pt}
\hrule height.#2pt}}}

\newcommand{\bl}{{\boldsymbol\cdot}}
\newcommand{\Conn}{W}
\newcommand{\Scal}{B}

\newcommand{\cScal}{\mathfrak{B}}
\newcommand{\cConn}{\mathcal{W}}

\newcommand{\cScalM}{\mathcal{P}}

\newcommand{\cL}{\mathcal{L}}
\newcommand{\re}{\mathbb{R}}

\newcommand{\Cont}{\widehat{W}}

\newcommand{\bd}{\boldsymbol d}
\newcommand{\cB}{\mathcal{B}}

\newcommand{\vx}{\vec{x}}
\newcommand{\vy}{\vec{y}}
\newcommand{\vu}{\vec{u}}
\newcommand{\dG}{\dot{G}_{(0,0)}}
\newcommand{\hsdG}{\dot{G}}

\newcommand{\cscal}{\mathfrak{b}}
\newcommand{\cscalm}{\mathfrak{p}}
\newcommand{\vp}{\vec{p}}
\newcommand{\vq}{\vec{q}}
\newcommand{\vv}{\vec{v}}
\newcommand{\udG}{G_{(0,0)}}

\newcommand{\umu}{\underline{\mu}}
\newcommand{\unu}{\underline{\nu}}
\newcommand{\vj}{\vec{j}}
\newcommand{\vpa}{\vec{\pa}}
\newcommand{\vh}{\varphi}

\begin{document}
\title{\thistitle}
\author{
	\kewang\textsuperscript{1},\ \leigh\textsuperscript{1},\ and\ \onkar\textsuperscript{1,2}\\
	\\	
{\small\textsuperscript{1}\emph{\addressuiuc}}\\
	\\
{\small\textsuperscript{2}\emph{\addresskitp}}\\
	\\
	}
\date{\today}
\maketitle\thispagestyle{empty}
\title

\begin{abstract}
We show that truncating the exact renormalization group equations of free $U(N)$ vector models in the single-trace sector to the linearized level reproduces the Fronsdal equations on $AdS_{d+1}$ for all higher spin fields, with the correct boundary conditions. More precisely, we establish canonical equivalence between the linearized RG equations and the familiar local, second order differential equations on $AdS_{d+1}$, namely the higher spin Fronsdal equations. This result is natural because the second-order bulk equations of motion on $AdS$ simply report the value of the quadratic Casimir of the corresponding conformal modules in the CFT. We thus see that the bulk Hamiltonian dynamics given by the boundary exact RG is in a different but equivalent canonical frame than that which is most natural from the bulk point of view.
\end{abstract}
\vspace{7cm}
\pagebreak

\parskip=10pt

\section{Introduction}
It is widely believed that gauge/gravity duality (holography) can be thought of as a geometrization of the renormalization group. In the most well-studied examples of holography, this has been investigated from the ``bulk to boundary'' point of view, where one starts with the bulk theory and deduces the renormalization group flow of the boundary quantum field theory by progressively integrating out the bulk geometry, but with no reference to any specific field theory cut-off. The first papers along these lines \cite{ deBoer:1999xf,deBoer:2000cz} noted the relationship between the boundary RG flow and Hamilton-Jacobi theory of the bulk radial evolution. Additional contributions were made for example by \cite{Alvarez:1998wr,Akhmedov:1998vf,Schmidhuber:1999rb,Henningson:1998gx,Balasubramanian:1999re,Skenderis:2002wp} and more recent discussions include \cite{Faulkner:2010jy,Heemskerk:2010hk,Gomes:2013qza} (see also \cite{Behr:2015yna}). However, a comprehensive understanding from the ``boundary to bulk'' perspective of the relationship between RG and holography still remains elusive, because of the lack of technology to control the RG flow of a generic, strongly coupled conformal field theory (see \cite{Lee:2009ij,Lee:2013dln} for some progress in this direction.\footnote{There have also been many other attempts at deriving holography from a boundary to bulk point of view. See for instance \cite{Gopakumar:2003ns, PhysRevD.86.065007, Nozaki:2012zj}.})

However, there exists a conjectured duality \cite{Klebanov:2002ja,Sezgin:2002rt,Leigh:2003gk} between free vector models in $d=2+1$ and certain types of higher-spin theories on $AdS_{4}$ constructed by Vasiliev (for a detailed exposition of higher-spin theories using Vasiliev's formalism, see for e.g., \cite{Vasiliev:1995dn, Vasiliev:2012vf, Vasiliev:1999ba, Bekaert:2005vh, Giombi:2009wh, Giombi:2012ms}). A detailed check of the proposal at the level of three-point functions was carried out in \cite{Giombi:2009wh, Giombi:2012ms} (see also \cite{Didenko:2012vh,Didenko:2012tv}). While the field theory side in this case is completely under control, the bulk is a far more complicated, highly non-linear (and non-local) theory involving fields of arbitrarily high spin propagating on $AdS$ space. Nevertheless, one might hope that this conjecture provides an accessible toy model for a ``constructive'' boundary to bulk understanding of holography. Indeed, using the collective-field formalism for free $O(N)$-vector models, it was demonstrated in a series of papers \cite{Das:2003vw, Koch:2010cy, Koch:2014mxa, Koch:2014aqa} that the boundary degrees of freedom can be reorganized to obtain the Fronsdal equation \cite{Fronsdal:1978rb,Fronsdal:1978vb}, to which the Vasiliev equations reduce at the linearized level (see \cite{Mintun:2014gua} for an RG-interpretation of this contruction). The conjectured vector model/higher spin duality therefore seems the natural playground to explore the relationship between holography and the renormalization group. 

Indeed, following the initial proposal of \cite{Douglas:2010rc} a holographic interpretation of the Wilson-Polchinksi exact renormalization group equations \cite{Polchinski:1983gv} for vector models sourced by single-trace operators was developed in \cite{Leigh:2014tza, Leigh:2014qca}. In the case of the $U(N)$-symmetric bosonic vector model for instance, the bulk theory consists of two bi-local fields $\cScal(z;\vx,\vy)$ and $\cScalM(z;\vx,\vy)$ evolving on a one-higher dimensional bulk space endowed with a flat connection.\footnote{Here $\vx,\vy\in \re^{1,d-1}$ are boundary spacetime events, and $z$ is the running RG scale, which is interpreted as the holographic radial direction.} By convention, the boundary is located at $z=\epsilon$, and the boundary values of $\cScal$ and $\cScalM$ are  the source and vacuum expectation value (VEV), respectively, for the single-trace bi-local operator $\mathcal{O}(\vx,\vy) = \phi^*_m(\vx)\phi^m(\vy)$ in the vector model. A point of great importance here is that $\cScal$ and $\cScalM$ coordinatize the bulk \emph{phase space}\footnote{The boundary generating functional of connected correlators of the single-trace operators is then the Hamilton-Jacobi functional for this system.}, and as such bulk dynamics is encoded in terms of radial evolution equations (namely Hamilton equations) for these fields. These equations of motion take the geometric form (see section \ref{prelim1} for details)
\beqn\label{rgi}
\mathcal{D}_z^{(0)} \cScal &=& \cScal\bl \Delta_B\bl \cScal
\\
\label{rgii}
\mathcal{D}_z^{(0)} \cScalM &=& iN\Delta_B -\cScalM\bl \cScal \bl \Delta_B-\Delta_B\bl \cScal\bl \cScalM
\eeqn
In \cite{Leigh:2014qca}, it was further shown that the bulk action evaluated on-shell organizes in terms of a Witten-diagram expansion, and precisely reproduces all the correlation functions of the boundary CFT.  One of the questions that was left unanswered in these papers was the emergence of the Fronsdal equations for individual higher spin fields from equations \eqref{rgi} and \eqref{rgii} at the linearized level -- showing this is the main goal of the present paper. We will demonstrate this cleanly in the case of odd boundary dimension, but we expect our arguments to also go through in even dimensions with slight modifications. 

There are two main obstacles to mapping the bulk equations derived from RG into Fronsdal equations: 

(i) Equation (\ref{rgi}) makes no reference to $\cScalM$ and thus can be solved by itself; being a first-order differential equation, how can the solutions look like general solutions of a second-order differential equation? 

(ii) The RG equations, even at the linearized order, are non-local (in the boundary directions); how can these be equivalent to local bulk equations? 

The resolution which will emerge below is as follows: $\cScal$ and $\cScalM$ are merely a particular choice of coordinates on the bulk phase space --- the one that the field theory gives us, and equations \eqref{rgi}, \eqref{rgii} are the corresponding Hamilton evolution equations. But we have the freedom to perform canonical transformations, without changing the physical content of the system. We will show that we can use this freedom to resurrect the $AdS_{d+1}$ Fronsdal equations in the bulk, i.e., \emph{there exists a canonical frame in which the linearized RG equations are precisely equivalent to local, second order differential equations in the bulk, namely the Fronsdal equations}. 

This result is in fact essentially guaranteed by group theory -- this is because of the fact that the Fronsdal equations simply express the quadratic Casimir \cite{Fronsdal:1978vb} of a particular lowest weight module of the conformal group $O(2,d)$. In fact, a given higher spin current in the field theory represents a specific conformal module of dimension $\Delta$ and spin\footnote{By spin $s$, we mean the irreducible traceless symmetric tensor representation with $s$ indices of $O(1,d-1)$.} $s$. Because the higher-spin currents are conserved at the free fixed point (or even at $N=\infty$ at the interacting fixed point), the module is short, with $\Delta=d-2+s$. The bi-local sources and vevs (i.e., the boundary 
 values of $\cScal$ and $\cScalM$) both consist of a direct sum of conformal modules
\beq\label{spinmodules}
\cScal(\epsilon;\vx,\vy) \in \oplus_{s=0}^{\infty} D(2-s,s),\;\;\;\cScalM(\epsilon;\vx,\vy) \in \oplus_{s=0}^{\infty}D(d-2+s,s)
\eeq
The holographic map is one-to-one between these boundary values and bulk fields, and thus the bulk fields fill out a reducible representation $\oplus_{s=0}^{\infty}\Big( D(2-s,s)\oplus D(d-2+s,s)\Big)$. The `diagonal' form of (\ref{spinmodules}) is present only at the boundary; the bulk dynamics mix together $\cScal(z;\vx,\vy)$ and $\cScalM(z;\vx,\vy)$. The challenge is then to find the right ``equivariant'' map between boundary RG and bulk dynamics (see \cite{Aizawa:2014yqa} for a recent discussion of $AdS$/CFT as an equivariant map between bulk and boundary representations). Interestingly, we can accomplish this simply with a canonical transformation!

This paper is organized as follows: in Section \ref{prelim1}, we review the construction of Refs. \cite{Leigh:2014tza,Leigh:2014qca}. In Section \ref{prelim2}, we review the form of the Fronsdal equations and express them in a gauge fixed form that is most useful from the point of view of lowest weight modules of $O(2,d)$. In Section \ref{section4}, we demonstrate explicitly the canonical transformation between the linearized renormalization group equations, and $AdS$ Fronsdal equations. We will end with some comments and general discussion in section \ref{section5}.

\section{Holography from the Renormalization group}\label{prelim1}

In order to be self-contained, we will review in this section some details of the holographic dual to the free bosonic $U(N)$ vector model constructed in \cite{Leigh:2014qca}.
\subsection{Free $U(N)$ vector model}
The action for the CFT is written in terms of $N$ complex scalars
\beq
S_0 = - \int d^d\vx\;\phi^*_m(\vx)\Box_{(\vx)}\phi^m(\vx)
\eeq
where $\vx^{\mu} \in \re^{1,d-1}$, $\Box_{(\vx)}=\eta^{\mu\nu}\vec{\pa}_{\mu}\vec{\pa}_{\nu}$, and $m=1,2\cdots N$ is the $U(N)$ index. The most general $U(N)$-invariant ``single-trace" deformations away from the free fixed point can be incorporated by introducing the two bi-local sources $B(\vx,\vy)$ and $W_{\mu}(\vx,\vy)$ as follows\footnote{It should be apparent that by choosing the sources to be of the \emph{quasi-local} form
\beq
B(\vx,\vy) = \sum_{s=0}^{\infty} B_{\mu_1\cdots\mu_s}(\vx)\;\vpa^{\mu_1}_{(x)}\cdots \vpa^{\mu_s}_{(x)}\delta^d(\vx-\vy)+\cdots
\eeq
\beq
W_{\mu}(\vx,\vy) = \sum_{s=0}^{\infty} W_{\mu;\mu_1\cdots\mu_s}(\vx)\;\vpa^{\mu_1}_{(x)}\cdots \vpa^{\mu_s}_{(x)}\delta^d(\vx-\vy)+\cdots
\eeq
we may source all the operators of interest, namely $\phi^*_m\phi^m,\;\phi^*_m\vec{\pa}^{\mu}\phi^m,\;\phi^*_m\vec{\pa}^{\mu}\vec{\pa}^{\nu}\phi^m\cdots$. Such operators are representative of specific conformal modules of spin $s$ and dimension $\Delta=d-2+s$.}  
\beq\label{action1}
S_{Bos.} = - \int_{\vx,\vu,\vy}\;\phi^*_m(\vx)\eta^{\mu\nu}D_{\mu}(\vx,\vu)D_{\nu}(\vu,\vy)\phi^m(\vy)+\int_{\vx,\vy}\phi_m^*(\vx)B(\vx,\vy)\phi^m(\vy)
\eeq
where we have defined (for reasons which will become clear shortly)
\beq
D_{\mu}(\vx,\vy) = P_{\mu}(\vx,\vy)+W_{\mu}(\vx,\vy),\;\;\;\;P_{\mu}(\vx,\vy) = \vec{\pa}_{\mu}^{(x)}\delta^d(\vx-\vy)
\eeq
Given the ``matrix'' notation we have introduced above, we can define a product and a trace between bi-locals as follows:
\beq
(f\bl g)(\vx,\vy) = \int_{\vu} f(\vx,\vu)g(\vu,\vy)
\eeq
\beq
\mathrm{Tr}\;(f) = \int_{\vx} f(\vx,\vx)
\eeq
We will largely use this notation from here on.

The sources $\Scal$ and $\Conn_{\mu}$ that we have introduced above couple, respectively, to the following bi-local operators
\beq
\hat{\Pi}(\vx,\vy) = \phi^*_m(\vy)\phi^m(\vx),\;\;\;\;\hat{\Pi}^{\mu}(\vx,\vy) = \int_{\vu}\Big(\phi^*_m(\vy)D^{\mu}(\vx,\vu)\phi^m(\vu)-D^{\mu}(\vy,\vu)\phi^*_m(\vu)\phi^m(\vx)\Big)
\eeq
Note that $\hat\Pi^{\mu}(\vx,\vy)$ can be interpreted as a bi-local current operator. There is an important subtlety in defining $U(N)$ singlet bi-local operators which should be pointed out -- since $\phi^m(\vx)$ is a section of a $U(N)$ vector bundle, the only natural contraction between $\phi^*_m(\vy)$ and $\phi^m(\vx)$ should involve a $U(N)$ Wilson line. For instance,
\beq
\hat\Pi(\vx,\vy) = \phi_m^*(\vy){\left(\mathscr{P}\;e^{\int_{\vy}^{\vx} A^{(0)}}\right)^m}_n\phi^n(\vx) \label{u(n)WL}
\eeq
where $A^{(0)}$ is a background $U(N)$ connection. In this paper, we will not include the Wilson lines explicitly; this is because we are assuming that the $U(N)$ vector bundle is trivial, which means that $A^{(0)}$ can be taken to be flat, and in particular we make the choice $A^{(0)}=0$.
 
The (unregulated) generating function (or partition function) is obtained by performing the path integral
\beq
Z_{CFT}[U,\Scal,\Conn] = \int \left[d\phi d\phi^*\right]e^{iU + iS_{Bos.}} \label{unregpf1}
\eeq
where we have introduced a new source $U$ (for the identity operator) to keep track of the overall normalization. 
\subsection{Background symmetries}
The path integration in \eqref{unregpf1} is over the set of all square integrable complex scalar functions over the space-time $\re^{1,d-1}$, namely $L_2(\re^{1,d-1})$. This space admits a natural action of ``unitary'' maps $\cL \in U(L_2(\re^{1,d-1}))$ (which we will henceforth refer to as $U(L_2)$ for convenience)\footnote{If we consider an infinitesimal version of the above transformation $\cL(\vx,\vy) \simeq \delta(\vx-\vy)+\epsilon(\vx,\vy)$, then the $U(L_2)$ condition implies $\epsilon^*(\vx,\vy)+\epsilon(\vy,\vx)=0$. For example, consider an $\epsilon$ of the form
$$\epsilon(\vx,\vy) = i\xi(\vx)\;\delta(\vx-\vy)+\xi^{\mu}(\vx)\;\vpa_{\mu}^{(x)}\delta(\vx-\vy)+i\xi^{\mu\nu}(\vx)\;\vpa_{\mu}^{(x)}\vpa_{\nu}^{(x)}\delta(\vx-\vy)+\cdots$$
where $\xi,\;\xi^{\mu},\;\xi^{\mu\nu}\cdots$ are all real. This satisfies the $U(L_2)$ condition provided $\vpa_{\mu}\xi^{\mu}=0,\;\vpa_{\mu}\xi^{\mu\nu}=0$ and so on. The first term above is an infinitesimal $U(1)$ gauge transformation, the second term is a volume-preserving diffeomorphism, while the rest are higher-derivative transformations.}
\beq
{\phi'}^m(\vx) = \int_{\vy}\cL(\vx,\vy)\phi^m(\vy),\;\;\cL^{\dagger}\bl \cL (\vx,\vy) = \delta^d(\vx-\vy)
\eeq
under which the path-integral measure is invariant. In fact, the path integral \eqref{pf1} has $U(L_2)$ as a \emph{background symmetry}, under which the sources $W_{\mu}$ and $B$ transform like a connection and an adjoint-valued field respectively:
\beq
W'_{\mu} = \cL^{-1}\bl W_{\mu} \bl \cL + \cL^{-1}\bl\left[P_{\mu},\cL\right]_{\bl},\;\;B'= \cL^{-1}\bl B\bl \cL
\eeq 
It was argued in \cite{Leigh:2014tza}, that the relevant geometry here is that of \emph{infinite jet bundles}, i.e., $\Conn_{\mu}$ is a connection 1-form on the infinite jet bundle over $\re^{1,d-1}$, while $B$ is a section of its endomorphism bundle. However, we will not need this language here. An important consequence of the above symmetry is that the free fixed point can be reached by setting $B=0$ and $W_{\mu} = W^{(0)}_{\mu}$, where $W^{(0)}_{\mu}$ is any flat connection 
\beq
dW^{(0)}+W^{(0)}\wedge W^{(0)} = 0
\eeq
where $d = dx^{\mu}\left[P_{\mu}, \cdot\right]_{\bl}$ is the exterior derivative. For this reason, we will find it convenient to pull out a flat piece from the full source $\Conn$ and write it as
\beq\label{connsplit}
\Conn = \Conn^{(0)}+\Cont
\eeq
Indeed, it is $\Cont$ and $\Scal$ which represent arbitrary single-trace, \emph{tensorial} deformations away from the free-fixed point, and thus parametrize single-trace RG flows away from the free CFT. 

In addition to $U(L_2)$, we also have a dilatation symmetry. In order to make this explicit, we introduce a conformal factor $z$ in the background metric $\eta_{\mu\nu}\mapsto z^{-2}\eta_{\mu\nu}$, and redefine the sources by rescaling them:
\beq
\Scal_{old} = z^{d+2} \Scal_{new},\;\;\;\Conn_{old}=z^d\Conn_{new}\label{new/old}
\eeq
For simplicity, we will drop the subscripts \emph{new} presently, and resurrect them when required. With these changes, the action takes the form
\beq
S_{Bos.}[\phi,z,\Scal,\Conn]=-\frac{1}{z^{d-2}}\int_{\vx,\vu,\vy}\phi_m^*(\vx)D_{\mu}(\vx,\vu)D^{\mu}(\vu,\vy)\phi^m(\vy)+\frac{1}{z^{d-2}}\int_{\vx,\vy}\phi_m^*(\vx)\Scal(\vx,\vy)\phi^m(\vy)\label{action2}
\eeq
It is clear now that the action is invariant under
\beq
\phi^m(\vx) \mapsto \lambda^{\frac{d-2}{2}}\;\phi^m(\vx),\qquad z \to \lambda\;z
\eeq
where $\lambda$ is a constant (i.e., spacetime independent) scale factor -- this larger symmetry group will be referred to as $CU(L_2)$. More generally, we could consider arbitrary Weyl transformations by making $\lambda$ $\vx$-dependent, but we will not do so here. Looking ahead, we note that $\mu \propto 1/z$ will end up being the effective ``renormalization scale''. Indeed, as we will see in the following section, the renormalization group flow will be parametrized by $z$. We will take $z$ to lie within the range $z\in [\epsilon,\infty)$, with $z=\epsilon$ corresponding to the ultraviolet, and $z\to \infty$ corresponding to the infra-red.

Finally, there is one major redundancy in our description which we need to fix -- given the tensorial nature of $\widehat{W}_{\mu}$, it is possible to set it to zero by redefining $B$ (as can be straightforwardly seen from equation \eqref{action2})
\beq
\cB = B -\left\{\Cont^{\mu},D^{(0)}_{\mu}\right\}_{\bl} -\Cont_{\mu}\bl\Cont^{\mu}
\eeq 
This is a special property of the bosonic theory.
We therefore arrive at the action
\beq
S_{Bos.} = - \frac{1}{z^{d-2}}\int_{\vx,\vu,\vy}\;\phi^*_m(\vx)\eta^{\mu\nu}D^{(0)}_{\mu}(\vx,\vu)D^{(0)}_{\nu}(\vu,\vy)\phi^m(\vy)+\frac{1}{z^{d-2}}\int_{\vx,\vy}\phi_m^*(\vx)\cB(\vx,\vy)\phi^m(\vy)
\eeq
where 
\beq
D^{(0)}_{\mu}(\vx,\vy) = P_{\mu}(\vx,\vy)+W^{(0)}_{\mu}(\vx,\vy)
\eeq 
We now move on to describe the renormalization group flow of the boundary field theory, and its holographic interpretation.

\subsection{Renormalization group as holography}
In order to study the renormalization group flow of the boundary field theory, we must regulate the path integral. Following Polchinski's formalism \cite{Polchinski:1983gv}, we will do so by introducing a smooth cutoff function $K_F(s)$ which has the property that $K_F(s) \to 1$ for $s < 1$ and $K_F(s) \to 0$ for $s >1$. 
We thus write the new action as
\beq
S^{reg.}_{Bos.} = - \frac{1}{z^{d-2}}\int_{\vx,\vy}\;\phi^*_m(\vx)K^{-1}_F\left(-z^2D^2_{(0)}/M^2\right)D^2_{(0)}(\vx,\vy)\phi^m(\vy)+\frac{1}{z^{d-2}}\int_{\vx,\vy}\phi_m^*(\vx)\cB(\vx,\vy)\phi^m(\vy)
\eeq
where $D_{(0)}^2 = \eta^{\mu\nu}D^{(0)}_{\mu}\bl D^{(0)}_{\nu}$, and $M$ is an auxiliary scale.\footnote{Note that this choice of regulator preserves the $U(L_2)$ symmetry. We used a slightly different regulator in \cite{Leigh:2014tza, Leigh:2014qca}. The present choice is somewhat more convenient -- the differences are merely notational, and not physical.} The particular choice of $K_F$ will not be important in our discussion below -- any sufficiently well-behaved cut-off function will do. In \cite{Mintun:2014gua}, the validity of such a cut-off procedure for the purposes of holography was questioned, because the cut-off is seemingly on the momenta of the fundamental $U(N)$ vectors, while only $U(N)$ invariant singlets should actually be visible in the bulk. However, we believe this objection is incorrect -- the point is that the cut-off procedure we utilize preserves the global $U(N)$ symmetry, and this is sufficient to ensure that the truncation to $U(N)$ invariant, single-trace operators is a consistent truncation. As we will see below, the RG equations are entirely written in terms of $U(N)$ singlet data, and $U(N)$ vectors are indeed invisible in the bulk. 

The regulated path integral is then given by
\beq
Z_{CFT}[z;U,\Scal,\Conn] = \int \left[d\phi d\phi^*\right]e^{iU + iS^{reg.}_{Bos.}} \label{pf1}
\eeq
It is clear that as we tune $z$ from $\epsilon$ to $\infty$, the effective cutoff for the field theory decreases from $\Lambda_{UV} = \frac{M}{\epsilon}$ to zero. In Wilsonian renormalization, this is interpreted as progressively integrating out fast modes. The partition function $Z$ must therefore remain unchanged under this process, and the effect of integrating out modes can be accounted for by making the source $\cB$ $z$-dependent. In this way, the exact RG equations are cast as exact Ward identities of the background $CU(L_2)$ symmetry. We will label the bulk  field $\cScal(z;\vx,\vy)$, to indicate that it lives in the one-higher dimensional bulk space. Similarly, the vev $\Pi(\vx,\vy)$ also evolves into a bulk field which we denote as $\cScalM(z;\vx,\vy)$. $\cScal$ and $\cScalM$ are in fact canonically conjugate fields which coordinatize the bulk phase space. Finally, along the RG trajectory, we also have the freedom to perform arbitrary $U(L_2)$ gauge transformations, and as a result, the connection $W^{(0)}$ also evolves into a flat connection in the bulk, which we label $\cConn_{(0)}$ (the $z$-component of which keeps track of the gauge transformations along the flow). The RG evolution equations are most conveniently obtained using Polchinski's formulation of the exact renormalization group \cite{Polchinski:1983gv}. We refer the reader to \cite{Leigh:2014qca} for a detailed derivation; we merely state the result here
\beq \label{rg1}
\mathcal{F}^{(0)} = \bd \cConn^{(0)}+\cConn^{(0)} \wedge_{\bl} \cConn^{(0)} = 0
\eeq
\beq\label{rg2}
\mathcal{D}^{(0)}_z\cScal  = \pa_z\cScal+\left[\cConn^{(0)}_z,\cScal\right]_{\bl}=\cScal\bl\Delta_B\bl\cScal
\eeq
\beq\label{rg3}
\mathcal{D}_z^{(0)} \cScalM = \pa_z\cScalM+\left[\cConn^{(0)}_z,\cScalM\right]_{\bl}=iN\Delta_B -\cScalM\bl \cScal \bl \Delta_B-\Delta_B\bl \cScal\bl \cScalM
\eeq
where $\boldsymbol{d}=dz\pa_z+d\vx^{\mu}\left[P_{\mu}, \cdot\right]_{\bl}$ is the bulk exterior derivative, and the bi-local kernel $\Delta_B$ is defined as
\beq
\Delta_B = \frac{2z}{M^2}\dot{K}_F\Big(-z^2D_{(0)}^2/M^2\Big)
\eeq
with $\dot{K}_{F}(s)=\pa_sK_F(s)$.\footnote{A convenient choice of regulator which we will use for computations in appendix B is the exponential cutoff $K_F(s)=e^{-s}$. In this case, the kernel $\Delta_B$ is proportional to the heat kernel for the operator $D_{(0)}^2$:
$\Delta_B = -\frac{2z}{M^2}e^{\frac{z^2}{M^2}D_{(0)}^2}$. In the limit $z\to 0$, $\Delta_B(z;\vx,\vy) \to -\frac{2z}{M^2}\delta^d(\vx-\vy)$. } We note that $\Delta_B$ defines a \emph{regulated} or \emph{smeared} version of the $\bl$-product between bi-locals (see footnote \ref{ftnt}).

Most importantly, the above equations are in fact the Hamilton equations of motion for the bulk Hamiltonian
\beq
\mathcal{H}_{bulk} = \mathrm{Tr}\;\cScalM\bl\Big(\left[\cScal,\cConn^{(0)}_z\right]_{\bl}+\cScal\bl\Delta_B\bl \cScal\Big)-iN\mathrm{Tr}\;\Delta_B\bl \cScal
\eeq
which in fact satisfies the \emph{Hamilton-Jacobi} relation
\beq
\mathcal{H}_{bulk} =-i\frac{\pa}{\pa z}\mathrm{ln}\;Z_{CFT}
\eeq
This is the central observation which leads to a holographic interpretation of the renormalization group equations with $z$ interpreted as the bulk radial coordinate. We can use the above Hamiltonian to construct a bulk action\footnote{\label{ftnt}The kernel $\Delta_B(z;\vx,\vy)$ defines a new regulated product, and a regulated trace between bilocals:
$$f*g(\vx,\vy) = \int_{\vu,\vv}f(\vx,\vu)\Delta_B(\vu,\vv)g(\vv,\vy)$$
$$\mathrm{Tr}_*(f) = \int_{\vx,\vy} \Delta_B(\vx,\vy)f(\vy,\vx)$$
If we define $\widetilde{\cScal} = \Delta_B^{-1}\bl \cScalM\bl \Delta_{B}^{-1}$, then the action takes the neater form
$$S_{bulk} = \int^{\epsilon}_{\infty}dz\;\mathrm{Tr}_*\left(\widetilde{\cScal}*\mathcal{D}_z^{(0)}\cScal-\widetilde{\cScal}*\cScal*\cScal\right)+iN\int^{\epsilon}_{\infty}dz\;\mathrm{Tr}_*(\cScal)$$
which is strikingly similar to non-commutative Chern-Simons action.}
\beq
S_{bulk}[\epsilon;\cScal,\cScalM] = \int^{\epsilon}_{\infty}dz\;\mathrm{Tr}\;\Big(\cScalM\bl \mathcal{D}_z^{(0)}\cScal - \cScalM\bl \cScal\bl\Delta_B\bl \cScal\Big)+iN\int^{\epsilon}_{\infty}dz\;\mathrm{Tr}\left(\Delta_B\bl \cScal\right)
\eeq
Solving the bulk equations of motion with the boundary conditions 
\beq
\cScal(\epsilon;\vx,\vy) = b^{(0)}(\vx,\vy),\qquad
\lim_{z\to\infty}\cScalM(z;\vx,\vy) = 0
\eeq
one can obtain the bulk action on-shell. It turns out that the on-shell action organizes itself in terms of a Witten-diagram expansion (see Fig. 1), and indeed, precisely reproduces the generating function of connected correlators in the boundary field theory \cite{Leigh:2014qca}
\beq
\boxed{Z_{CFT}[\epsilon; b^{(0)}] = e^{iS^{o.s}_{bulk}[\epsilon, b^{(0)}]}}
\eeq 
which is of course the statement of holographic duality.\begin{figure}[!h]\label{fig1}
\centering
\begin{tabular}{l l l l l}
\includegraphics[height=4cm, width=3.8cm]{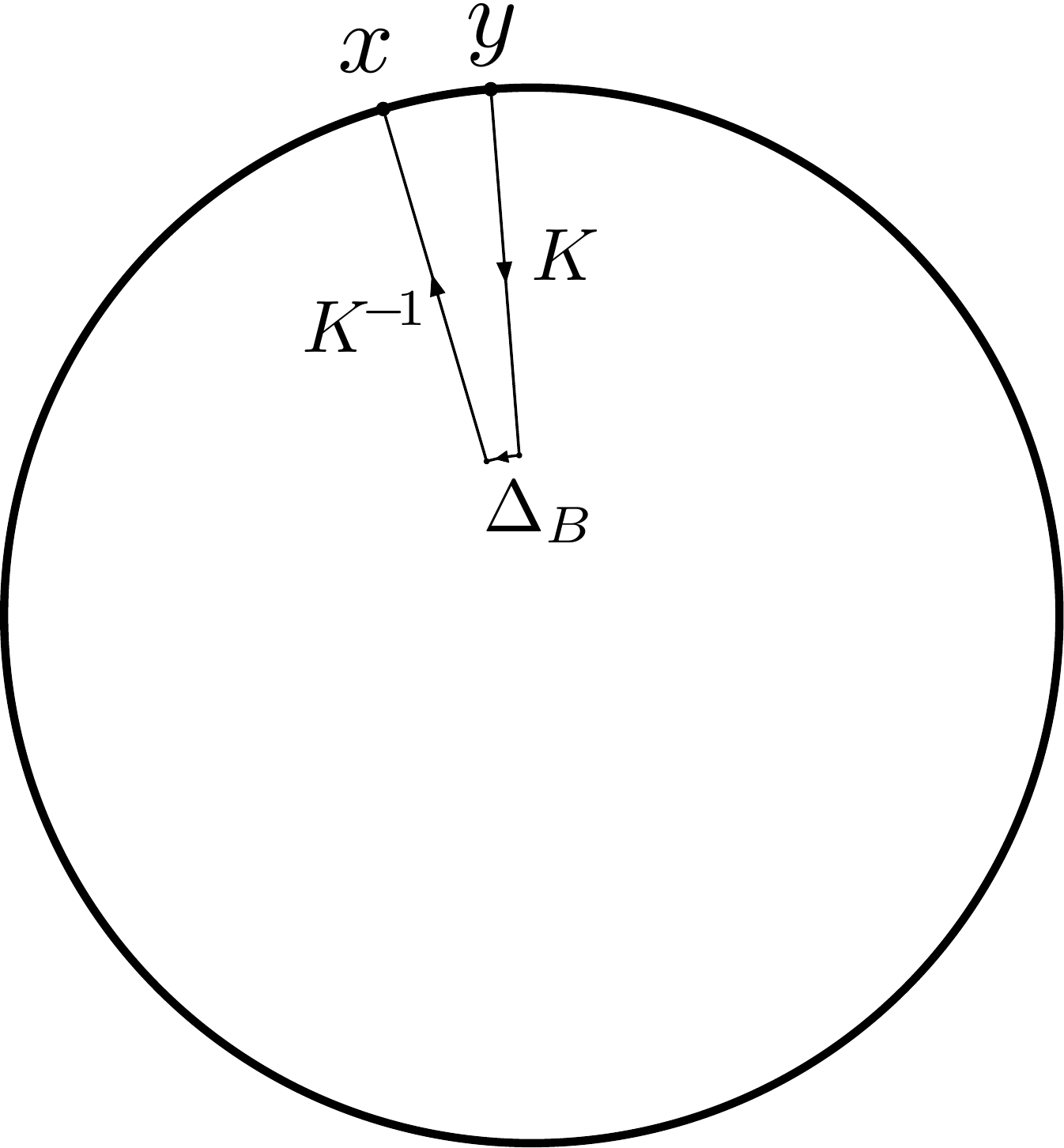} & &\includegraphics[height=4cm, width=3.8cm]{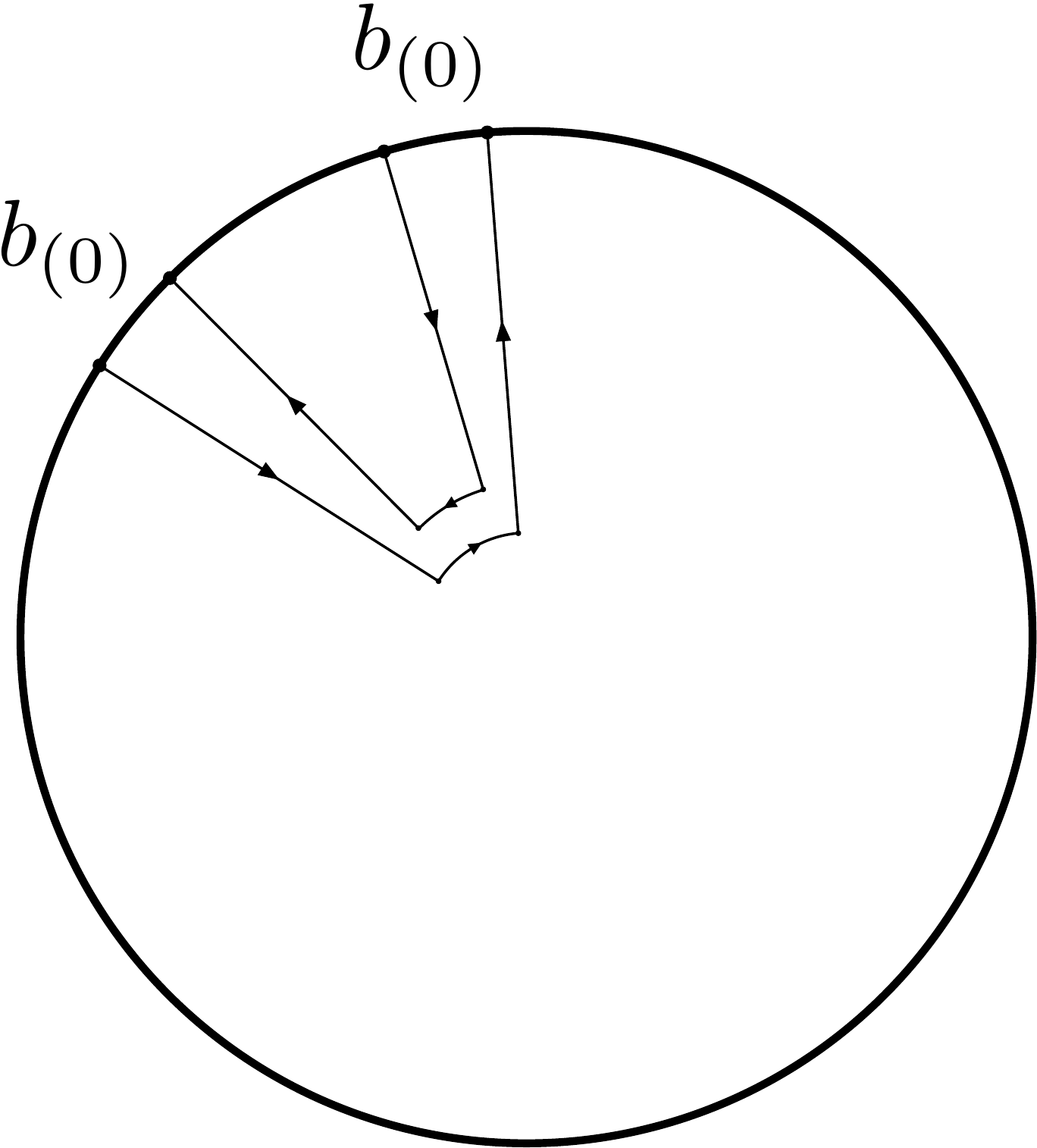} & &\includegraphics[height=4cm, width=4cm]{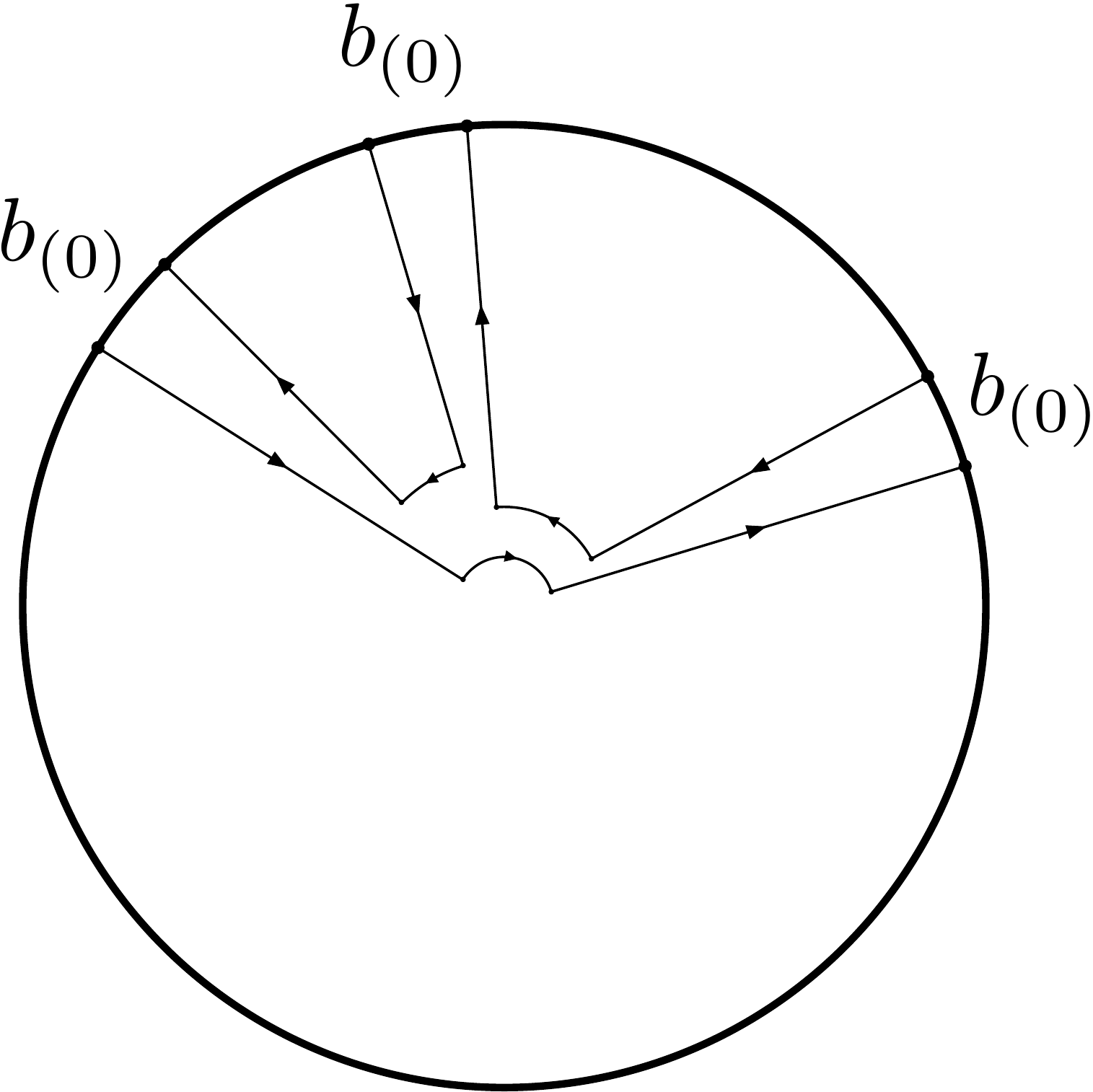}
\end{tabular}
\caption{\small{The holographic Witten diagrams corresponding to the boundary one, two and three point functions. The radial lines are in/out-going Wilson lines for the flat connection $\cConn^{(0)}$, while horizontal lines correspond to $\Delta_B$. Note that the boundary loops are ``pulled-in'' into bulk vertices.}}
\end{figure}

One must note that the bulk theory contains all multi-point interactions, and these are non-local. Of course the reason for this is that we have organized an infinite number of massless fields in the bulk into a bi-local form. Any theory containing an infinite number of massless fields should be thought of as non-local and so this should come as no surprise. 
The aim of the rest of the paper will be to isolate the propagating fields of fixed spin; the first such task will be to reproduce the Fronsdal equations in $AdS$ space from equations (\ref{rg2}, \ref{rg3}). Before proceeding, we present a brief overview of Fronsdal equations in $AdS$ space.

\section{The Fronsdal equation}\label{prelim2}
The Fronsdal higher spin theory in $AdS$ space is described by symmetric tensors $\vh_{I_1 ... I_n}$ which satisfy the double-tracelessness conditions $\vh'' {}_{I_5 ... I_n}\equiv g^{I_1I_2}g^{I_3I_4}\vh_{I_1...I_n}=0$. Here the bulk  coordinate indices run over the boundary coordinate indices $\mu=0,1,...,d-1$ and the radial direction $z$, i.e., $I=(\mu,z)$. The equations of motion are explicitly
\begin{eqnarray}
\nabla_I \nabla^I \vh_{I_1...I_n} - n \nabla_I \nabla_{(I_1} \vh_{I_2 ... I_n)}{}^I
+\tfrac{1}{2}n(n-1) \nabla_{(I_1} \nabla_{I_2} \vh_{I_3...I_n)I}{}^I-2(n-1)(n+d-2) \vh_{I_1 ... I_n} = 0 
\label{fronsdal}
\end{eqnarray}
where the indices $I_1,\cdots I_n$ should be taken to be symmetrized as indicated by parentheses. These equations are invariant under the gauge transformations
\begin{equation}
\delta_\Lambda \vh_{I_1 ... I_n} = \nabla_{(I_1} \Lambda_{I_2 ... I_n)}
\end{equation}
where $\nabla$ is the $AdS_{d+1}$ covariant derivative and the symmetric gauge parameters $\Lambda_{I_2 ... I_n}$ satisfy the single-tracelessness conditions $g^{I_2I_3}\Lambda_{I_2 ... I_n}\equiv \Lambda' {}_{I_4 ... I_n} = 0$. For $n=1$, equation \eqref{fronsdal} is the familiar Maxwell's equation, while for $n=2$ it is the linearized Einstein's equation. 

Such a presentation of the higher spin equations is inconvenient in the present context. We wish to isolate specific (lowest weight) representations of $O(2,d)$; such representations are given by irreducible spin-$s$ representations of $SO(1,d-1)$. We can accomplish this by appropriately fixing the gauge invariance. Many different choices of gauge have been considered in the literature, but the appropriate one here is the ``Coulomb gauge''\footnote{The terminology ``gauge'' is somewhat incorrect in this context -- what is being said really, is that the fields with $z$-indices $\vh_{z\cdots z \mu_1\cdots \mu_k}$ are non-dynamical, in the sense that they do not contribute to the symplectic structure.}
\beq
\vh_{\underbrace{z...z}_{m}\mu_1...\mu_s}=0,\qquad \pa^\mu \vh_{\mu \mu_1 ...\mu_s}=0\qquad \forall m>0, \forall s
\eeq
In addition, in order to have an irreducible $SO(1,d-1)$ representation, we require $\vh^{\mu}{}_{\mu \mu_1 ...\mu_{s-2}}=0$.
In this gauge, the equations of motion reduce to 
\beq\label{Coulombeom}
\left[z^2\pa_z^2+(2s-d+1)z\pa_z+s(s-d)+(2-s)(s+d-2)+z^2\Box_{(\vx)}\right] \vh_{\mu_1...\mu_s}(z,\vx)=0
\eeq
where $\Box_{(\vx)} = \eta^{\mu\nu}\vpa_{\mu}\vpa_{\nu}$. In Appendix \ref{app:gauge}, we will discuss this gauge in some more detail. However, it is illuminating to obtain equation (\ref{Coulombeom}) directly from the $AdS$/CFT point of view, as a statement about the matching of quadratic Casimirs between the bulk and boundary representations \cite{Fronsdal:1978vb}. Starting from the CFT, consider a local, symmetric, traceless, spin $s$, quasi-primary operator $\hat{\cal O}_{a_1...a_s}(0)$ of dimension $\Delta$ in the boundary CFT (where $a_k = 0, \cdots d-1$ are boundary indices). Such an operator satisfies (by definition)
\beqn\label{CFTop}
\left[K_a, \hat{\cal O}_{a_1...a_s}(0)\right]&=&0\\
\left[M_{ab}, \hat{\cal O}_{a_1...a_s}(0)\right]&=&
\Sigma_{ab}(\hat{\cal O}_{a_1...a_s}(0))=-is \eta_{a(a_1}\hat{\cal O}_{a_2...a_s)b}(0)+is \eta_{b(a_1}\hat{\cal O}_{a_2...a_s)a}(0) \\
\left[D, \hat{\cal O}_{a_1...a_s}(0)\right]&=&-i\Delta \hat{\cal O}_{a_1...a_s}(0)
\eeqn
where $\Sigma_{ab}$ is the appropriate spin matrix. The quadratic Casimir of the conformal group is given by
\beq\label{Casimir}
C_2^{O(2,d)} = -D^2+\frac{1}{2}M_{ab}M^{ab}-\frac{1}{2}\left\{P_{a},K^{a}\right\}
\eeq
From equations \eqref{CFTop} and \eqref{Casimir}, we find straightforwardly\footnote{This result is independent of the spacetime location $\vx$ of the operator, because the quadratic Casimir commutes with translations. Equivalently, every element of the conformal module of course shares the same value of the Casimir.}
\beq
\left[C_2^{O(2,d)},\hat{\cal O}_{a_1...a_s}(\vx)\right]=\Big(-\Delta(d-\Delta)+s(s+d-2)\Big)\hat{\cal O}_{a_1...a_s}(\vx)
\eeq

Now the corresponding bulk field $\vh_{a_1\cdots a_s}$ of course must have the same value for the Casimir, as it transforms in the corresponding dual $AdS$ representation. We note that $O(2,d)$ is represented in the bulk as
\beqn
\left[D, \varphi_{a_1...a_s}(z,\vx)\right] 
&=&i\vx^a\left[P_a, \varphi_{a_1...a_s}(z,\vx)\right]
+iz\pa_z\varphi_{a_1...a_s}(z,\vx)
\nonumber\\
\left[M_{ab},\varphi_{a_1...a_s}(z,\vx)\right]
&=&i\vx_a\left[P_b, \varphi_{a_1...a_s}(z,\vx)\right]
-i\vx_b\left[P_a, \varphi_{a_1...a_s}(z,\vx)\right]
+\Sigma_{ab}(\varphi_{a_1...a_s})(z,\vx)
\nonumber\\
\left[K_a, \varphi_{a_1...a_s}(z,\vx)\right] 
&=&
 -i(2\vx_a \vx^b-(\vx^2+z^2)\delta^b_a)\left[P_b , \varphi_{a_1...a_s}(z,\vx)\right]
-i2\vx_az\pa_z\varphi_{a_1...a_s}(z,\vx)\nonumber\\
&-&2\vx^b\Sigma_{ab}(\varphi_{a_1...a_s})(z,\vx)
\nonumber\\
\left[P_a,\varphi_{a_1...a_s}(z,\vx)\right]&=&i\vpa_a\varphi_{a_1...a_s}(z,\vx)\nonumber
\eeqn
In this bulk representation, we then have
\beqn
\left[C_2^{O(2,d)},\varphi_{a_1...a_s}(z,\vx)\right]&=& z^2\pa^2_z\varphi_{a_1...a_s}(z,\vx))+(-d+1)z\pa_z\varphi_{a_1...a_s}(z,\vx)\nonumber\\&&
+s(s+d-2)\varphi_{a_1...a_s}(z,\vx)-z^2\left[P^a,\left[P_a,\varphi_{a_1...a_s}(z,\vx)\right]\right]
\eeqn
But from the CFT calculation, we know that $C_2^{O(2,d)}=-\Delta(d-\Delta)+s(s+d-2)$; therefore, requiring that the two Casimirs agree gives us
\beq
 z^2\pa^2_z\varphi_{a_1...a_s}(z,\vx)+(-d+1)z\pa_z\varphi_{a_1...a_s}(z,\vx)
+\Delta(d-\Delta)\varphi_{a_1...a_s}(z,\vx)+z^2\left[P^a,\left[P_a,\varphi_{a_1...a_s}(z,\vx)\right]\right]=0
\eeq
To compare this with equation \eqref{Coulombeom}, we simply note that in the bulk representation, 
the $a,b,...$ indices must be interpreted as those corresponding to a local frame, as it is in that case that $O(1,d-1)$ acts in the simple fashion stated. Converting to coordinate indices, $\varphi_{a_1...a_s}(z,\vx)$ becomes $z^s \vh_{\mu_1...\mu_s}(z,\vx)$. Inserting this in the above equation gives 
\beq
\left[ z^2\pa^2_z +(2s-d+1)z\pa_z
+s(s-d)+\Delta(d-\Delta)+z^2\Box_{(\vx)}\right] \vh_{\mu_1...\mu_s}(z,\vx)=0
\eeq
In the case when the boundary operator is in a short representation, i.e., $\hat{\cal O}_{a_1\cdots a_s}$ is a conserved current, we have $\Delta=s+d-2$, and so this becomes
\beq
\left[ z^2\pa^2_z +(2s-d+1)z\pa_z
+s(s-d)+(2-s)(s+d-2)+z^2\Box_{(\vx)}\right] \vh_{\mu_1...\mu_s}(z,\vx)=0
\eeq
in agreement with (\ref{Coulombeom}). So we conclude that indeed the linearized higher spin equations simply state the value of the Casimir of the appropriate conformal module. Consequently, it must be that eqs. (\ref{rg2}-\ref{rg3}) should yield the Fronsdal equations. In the rest of the paper, we proceed to show this explicitly.

\section{From Wilson-Polchinski to Fronsdal}\label{section4}
Let us now embark on our main goal of reproducing the $AdS$-Fronsdal equations from the Wilson-Polchinksi exact renormalization group equations:
\beq\label{rgI}
\mathcal{D}_z^{(0)} \cScal = \cScal\bl \Delta_B\bl \cScal
\eeq
\beq\label{rgII}
\mathcal{D}_z^{(0)} \cScalM = iN\Delta_B -\cScalM\bl \cScal \bl \Delta_B-\Delta_B\bl \cScal\bl \cScalM
\eeq
In particular, we want to study the above equations upon linearizing about the background 
\beq
\cScal = 0,\;\;\;\cScalM = \cScalM^{(0)}
\eeq
where $\cScalM^{(0)}$ satisfies $\mathcal{D}_z^{(0)}\cScalM^{(0)} = iN\Delta_B$. Clearly, this background is a solution of the equations \eqref{rgI} and \eqref{rgII}, albeit the trivial one which corresponds to the unperturbed boundary CFT. We introduce an auxiliary expansion parameter $\lambda$ and write\footnote{This is where large $N$ plays an important role because such an expansion exists in practice only at large $N$, with $1/N$ providing the expansion parameter.} 
\beq
\cScal(z;\vx,\vy) = \lambda\;\mathfrak{b}_1(z;\vx,\vy)+O(\lambda^2),\;\;\cScalM(z;\vx,\vy) = \cScalM^{(0)}(z;\vx,\vy) + \lambda\;\mathfrak{p}_1(z;\vx,\vy)+O(\lambda^2)
\eeq
At linear order in $\lambda$, we thus obtain the equations
\beqn
\mathcal{D}_z^{(0)} \cscal_1 &=& 0
\\
\mathcal{D}_z^{(0)} \cscalm_1 &=& -\mathcal{P}^{(0)}\bl \cscal_1\bl\Delta_B-\Delta_B\bl \cscal_1\bl \mathcal{P}^{(0)}
\eeqn
Also recall, that these equations were written for the ``$new$'' fields defined in \eqref{new/old}. We now revert back to the ``$old$'' fields by restoring the appropriate powers of $z$: 
$$\cscal^{new}_1 = \frac{1}{z^{d+2}}\cscal^{old}_1,\;\;\cscalm^{new}_1 = \frac{1}{z^{d-2}}\cscalm_1^{old}$$
With this replacement, we get 
\beqn
\mathcal{D}_z^{(0)} \cscal^{old}_1&=& \frac{(d+2)}{z}\;\cscal^{old}_1\label{EOMlin1}
\\
\mathcal{D}_z^{(0)} \cscalm^{old}_1 &=& \frac{d-2}{z}\cscalm^{old}_1-\frac{1}{z^{4}}\left(\mathcal{P}^{(0)}\bl \cscal^{old}_1\bl\Delta_B+\Delta_B\bl \cscal^{old}_1\bl \mathcal{P}^{(0)}\right) \label{EOMlin2}
\eeqn
In the rest of the paper, we will restrict our attention to the case of odd boundary dimension $d$, with brief comments about even $d$ towards the end. 
\subsection{Spin-zero}
For simplicity, let us practice with the spin $s=0$ case first, before moving on to the arbitrary spin case. In other words, we turn on bulk fields which are dual to the $s=0$ operator $J^{(0)}(\vx)=:\phi^*_m\phi^m:(\vx)$ in the boundary field theory. To that effect, we take\footnote{Here the bulk field $\phi(z,\vx)$ should not be confused with the elementary scalar $\phi^m(\vx)$ of the boundary field theory.} 
\beq
\cscal^{old}_1(z;\vx,\vy) = \phi(z,\vx)\Big(z^d\delta^d(\vx-\vy)\Big) \label{lin1}
\eeq
\beq
\pi(z,\vx) = \frac{1}{N}\lim_{\vx\to \vy}\cscalm^{old}_1(z;\vx,\vy) = \frac{1}{N}\langle J^{(0)}\rangle_1(z,\vx) \label{lin2}
\eeq 
The above projection onto local fields is consistent only because we are working at the linearized level, where the different spins are decoupled in the bulk (as we will see explicitly below). Note that the operator $J^{(0)}(\vx)$ above is ``normal ordered'' with respect to the free CFT, meaning 
\beq
J^{(0)}(\vx) = \lim_{\vy\to\vx}\Big(\phi^*_m(\vx)\phi^m(\vy)-\left\langle \phi^*_m(\vx)\phi^m(\vy)\right\rangle_{CFT}\Big)
\eeq
and the subscript $\langle J^{(0)}\rangle _1$ in equation \eqref{lin2} stands for linearized order in $\alpha$. The linearized equations of motion \eqref{EOMlin1}, \eqref{EOMlin2} become
\beq
z\pa_z\phi(z,\vx) = \Delta_-\;\phi(z,\vx)\label{cs1}
\eeq
\beq
z\pa_z\pi(z,\vx) = \Delta_+\pi(z,\vx)-z^{2\nu+1}\int_{\vu}\frac{1}{N}\Big(\mathcal{P}^{(0)}(z;\vx,\vu)\Delta_B(z;\vu,\vx)+\Delta_B(z;\vx,\vu) \mathcal{P}^{(0)}(z;\vu,\vx)\Big)\phi(z,\vu)\label{linRG0}
\eeq
where we have defined 
\beq
\Delta_+ = d-2,\qquad\Delta_- = 2,\qquad\Delta_+-\Delta_- = 2\nu
\eeq
To simplify the notation somewhat, we rewrite the above equations in the compact form
\beqn
z\pa_z\phi(z,\vx) &=& \Delta_-\;\phi(z,\vx) \label{linRG1}\\
z\pa_z\pi(z,\vx) &=& \Delta_+\pi(z,\vx)+\frac{z^{2\nu}}{2}\int d^d\vu\;\dot{G}_{(0,0)}(z;\vx,\vu)\phi(z,\vu)\label{linRG2}
\eeqn
where the meaning of $\dot{G}_{(0,0)}$ will become clear shortly. These equations of motion come from the linearized action
\beq
S^{(2)}_{bulk} = \int^{\infty}_{\epsilon} \frac{dz d^d\vx}{z^{d+1}}\;\left(\pi(z,\vx) z\pa_z\phi(z,\vx)-\Delta_-\pi(z,\vx)\phi(z,\vx) +\int d^d\vy\;\frac{z^{2\nu}}{4}\phi(z,\vx)\dot{G}_{(0,0)}(z;\vx,\vy) \phi(z,\vy)\right)\label{spin0action}
\eeq
A convenient way to keep track of the boundary condition on $\phi(z,\vx)$ at $z=\epsilon$ is to add the boundary term
\beq\label{bdryspin0action}
S_{bdry} = \frac{1}{\epsilon^d}\int d^d\vx\;\pi(\epsilon,\vx)\left(\phi(\epsilon,\vx)-\epsilon^{\Delta_-}\phi^{(0)}(\vx)\right)
\eeq
to the action. Our aim now is to show that equations \eqref{linRG1}, \eqref{linRG2} are completely equivalent to the Fronsdal equation for spin $s=0$. 

As mentioned in the introduction, there are two main obstacles we must confront: (i) a confusing property of the above action (and the corresponding Hamiltonian) is the absence of a $\pi^2$ term. Naively, this gives the impression of a lack of any interesting dynamics. Another manifestation of this problem is that the field $\phi$ seems to satisfy an ultra-local first order equation, which is obviously not true of the usual bulk fields in $AdS$/CFT. (ii) The other problem is that the $\pi$ equation of motion seems non-local, due to the presence of the bilocal kernel $\dot{G}_{(0,0)}$. 

To resolve these issues, we must remember that we're in a phase space formulation -- $\phi$ and $\pi$ are coordinates on the bulk phase space, with the symplectic structure\footnote{We use bold symbols $\boldsymbol{\delta\phi}$, $\boldsymbol{\delta\pi}$ etc. to denote differential 1-forms on the phase space.}
\beq
\boldsymbol{\Omega}(z) = \int \frac{d^d\vx}{z^d}\;\boldsymbol{\delta\phi}(z,\vx)\wedge \boldsymbol{\delta\pi}(z,\vx)
\eeq
In the specific symplectic frame coordinatized by $\phi$ and $\pi$,  $\phi(z,\vx)$ is fixed through (\ref{linRG1}) by its boundary value, and $\pi(z,\vx)$ contains all of the information about the renormalized 2-point function of the current. Indeed, it is straightforward to see from equations \eqref{linRG0} and \eqref{linRG2} that if we define
\beq
G_{(0,0)}(z;\vx,\vy) = \frac{2i}{N}\left\langle J^{(0)}(\vx)J^{(0)}(\vy)\right\rangle_{CFT,Mink}(z) \label{defG}
\eeq 
then
\beq
\dot{G}_{(0,0)}(z;\vx,\vy) = \frac{2i}{N}z\pa_z\left\langle J^{(0)}(\vx)J^{(0)}(\vy)\right\rangle_{CFT,Mink}(z),
\eeq 
where the correlator is defined in the \emph{regulated} CFT on Minkowski space, with the cut-off procedure described in section 2 (see appendix B.1 for more details). 

An essential feature of the phase space formulation is that we have the freedom to perform \emph{canonical (symplectic) transformations}, which are field redefinitions (i.e., coordinate transformations on phase space) which leave the symplectic 2-form unchanged. Consider for instance, a general linear transformation on phase space
\beqn
\phi &=& A\bl \varphi+B\bl\varpi\nonumber\\
\pi &=& C\bl\varphi+D\bl\varpi
\eeqn
for general bi-local kernels $A, B, C, D$. The requirement that the symplectic 2-form be preserved, namely
\beq
\int\frac{d^d\vx}{z^d}\; \boldsymbol{\delta\phi}(z,\vx)\wedge \boldsymbol{\delta\pi}(z,\vx)= \int\frac{d^d\vx}{z^d}\;\boldsymbol{\delta \varphi}(z,\vx)\wedge \boldsymbol{\delta\varpi}(z,\vx) 
\eeq
leads to the constraints
\beq
A^T\bl C = C^T\bl A,\;\;D^T\bl B = B^T\bl D \label{SC1}
\eeq
\beq
A^T\bl D - C^T\bl B = \boldsymbol{1}\label{SC2}
\eeq
For simplicity (and because this suffices for our purpose), we will restrict our attention to the case where $A, B, C, D$ are symmetric, and translationally and rotationally invariant. In this case, the constraints \eqref{SC1} are automatically satisfied, and we only have to satisfy the constraint \eqref{SC2}. 

To avoid unnecessary complications, we begin by choosing a simpler canonical transformation\footnote{A similar transformation also appeared in \cite{Lee:2013dln}, although higher-derivative corrections were not under control in that case. We also note that in the quantum RG formulation of \cite{Lee:2013dln}, canonical transformations are simply changes of integration variables in the bulk path-integral, which leave the measure invariant.}
\beqn
\phi(z,\vx) &=& \varphi(z,\vx)+ \frac{2\delta }{z^{2\nu}}\int_{\vy}\dG^{-1}(z;\vx,\vy)\varpi(z,\vy)\nonumber\\
\pi(z,\vx) &=&  \varpi(z,\vx)\label{CT1}
\eeqn
for some constant $\delta$ to be fixed later. This ansatz clearly satisfies all of the constraints (because $\dG^{-1}$ is a symmetric kernel), and is therefore a canonical transformation. We will presently show that for a specific choice of $\delta$, the field $\varphi$ satisfies the spin-zero $AdS_{d+1}$ Fronsdal equation, up to higher-derivative corrections (i.e., up to $O(z^4\vpa^4)$ terms). We will later show that these higher derivative terms can in fact be systematically eliminated by a more sophisticated choice of the canonical transformation, but we postpone that discussion to section \ref{3.3}. 

Substituting equation \eqref{CT1} into \eqref{spin0action}, the action in terms of the new fields becomes
\beqn
S^{(2)}_{bulk} &=& \int \frac{dz}{z}\;\Big( \frac{1}{z^d}\varpi \bl \left(z\pa_z\varphi-(\Delta_--\delta)\varphi\right) 
+ \frac{2\delta}{z^d}\varpi \bl z\pa_z\left(z^{-2\nu}\dG^{-1}\bl\varpi\right) \nonumber\\
&-&\frac{(2\Delta_--\delta)\delta}{z^{d+2\nu}}\varpi\bl\dG^{-1}\bl\varpi+\frac{1}{4z^{d-2\nu}}\varphi\bl \dG\bl \varphi\Big)
\eeqn 
where we have switched to the $\bl$-product notation for convenience. Let us focus on the second term above:
\beqn
2nd\;\mathrm{term}&=& -4\nu\delta\int \frac{dz}{z}\frac{1}{z^{d+2\nu}}\varpi\bl \dG^{-1}\bl \varpi+2\delta\int \frac{dz}{z}\frac{1}{z^{d+2\nu}}\varpi\bl z\pa_z\left(\dG^{-1}\bl \varpi\right)\nonumber\\
&=& -4\nu\delta\int \frac{dz}{z}\frac{1}{z^{d+2\nu}}\varpi\bl \dG^{-1}\bl \varpi+2\delta\int dz\frac{1}{z^{d+2\nu}}\varpi\bl\left( \pa_z(\dG^{-1})\bl \varpi+\dG^{-1}\bl \pa_z\varpi\right)\nonumber\\
& = & \delta(d-2\nu)\int \frac{dz}{z}\frac{1}{z^{d+2\nu}}\varpi\bl \dG^{-1}\bl \varpi+\delta\int \frac{dz}{z}\frac{1}{z^{d+2\nu}}\varpi\bl z\pa_z(\dG^{-1})\bl \varpi\nonumber\\
&-&\frac{\delta}{\epsilon^{d+2\nu}}\left.\varpi\bl \dG^{-1}\bl \varpi\right|_{z=\epsilon}
\eeqn
where in the last line we have integrated by parts with respect to $z$. Putting everything together, we get the bulk action
\beq
S^{(2)}_{bulk} = \int \frac{dz}{z^{d+1}}\;\Big(\varpi \bl z\pa_z\varphi -(\Delta_--\delta)\varpi\bl\varphi+ \frac{1}{z^{2\nu}}\varpi\bl\left[\delta^2\dG^{-1}+\delta z\pa_z(\dG^{-1})\right]\bl\varpi+\frac{1}{4z^{-2\nu}}\varphi\bl \dG\bl \varphi\Big)\label{spin0action2}
\eeq 
Evidently, the new action has a $\varpi^2$ term in it, as opposed to the previous version. Of course, the integration by parts we have performed above also produces a new boundary term
\beq\label{bdryspin0action2}
\delta S_{bdry} = -\frac{\delta}{z^{d+2\nu}}\left.\varpi\bl \dG^{-1}\bl \varpi\right|_{z=\epsilon}
\eeq
This boundary term has a clear interpretation from the bulk point of view -- it is the \emph{generating function} for the canonical transformation. From the boundary point of view, it appears to be a multi-trace deformation. We will return to the boundary terms shortly. 

In order to proceed, we need to examine the various bi-local kernels appearing in the above equations. The kernel $\dot{G}_{(0,0)}$ admits an asymptotic expansion of the form (see Appendix \ref{appB})
\beq\label{G}
\dG(z;\vx,\vy)=-\frac{C}{z^{2\nu}}\left(1+\alpha z^2\Box_{(\vx)}+\cdots\right)\delta^d(\vx-\vy)
\eeq
\beq\label{Ginv}
\dG^{-1}(z;\vx,\vy) = -\frac{z^{2\nu}}{C}\left(1-\alpha z^2\Box_{(\vx)}+\cdots\right)\delta^d(\vx-\vy)
\eeq
where $\alpha>0$ and $C$ are (dimensionful) constants, which are evaluated in the Appendix. While the numerical values of these constants are irrelevant, the positivity of $\alpha$ is important in the present discussion for the bulk metric to have the correct signature. The ellipsis above indicate higher-derivative terms, which we will address in Section \ref{3.3}, because presently our aim is to obtain a two-derivative action. An intuitive way to understand the above expansions is as follows: in any CFT, the two point function of a given operator is universally determined by conformal invariance. Ambiguities which arise upon introducing a regulator come in the form of \emph{local counterterms} -- equations \eqref{G} and \eqref{Ginv} parametrize precisely such counterterms.  

The $\varpi^2$ term in the action simplifies to
\beq
\frac{1}{C}\int \frac{dzd^d\vx}{z^{d+1}}\varpi(z,\vx)\Big(-\delta(\delta+2\nu)+\alpha \delta(\delta+2\nu+2) z^2\Box_{(\vx)}+\cdots\Big)\varpi(z,\vx)
\eeq
where again the ellipsis indicates higher-derivative terms. To see that the action \eqref{spin0action2} gives rise to the spin-zero $AdS_{d+1}$-Fronsdal equation, we write down the equations of motion:
\beqn
z\pa_z\varphi-(\Delta_--\delta)\varphi &=&
\frac{2\delta}{C}\Big((2\nu+\delta)\varpi-\alpha(2\nu+\delta+2)z^2\Box_{\vx}\varpi+...\Big)\label{psfr1}\\
-z\pa_z\varpi +(\Delta_++\delta)\varpi &=& 
\frac{C}{2}\Big(\varphi+\alpha z^2\Box_{(\vx)}\varphi+\cdots\Big)\label{psfr2}
\eeqn
Combining these two equations into a second order differential equation, we get (up to $O(z^4\vpa^4)$ terms)
\beqn
z\pa_z(z\pa_z\varphi)-dz\pa_z\varphi
+\Delta_-\Delta_+\varphi
-2\alpha\delta z^2\Box_{(\vx)}\varphi
=-\frac{4\alpha\delta (2\nu+\delta+2)}{C}z^2\Box_{\vx}\varpi
+\cdots\label{linscalareom2}
\eeqn
We see that the right-hand side of (\ref{linscalareom2}) can be removed (and thus is of order $z^4\vpa^4$) with the choice $\delta=-(2\nu+2)$. (Equivalently, the second term on the right hand side of \eqref{psfr1} drops out with this choice of $\delta$.) Further, by rescaling the $\vx$ coordinates, we can set the coefficient $-2\alpha\delta=2\alpha(2\nu+2)>0$ of the $\Box_{(\vx)}$ term to one. We thus recognize the above equation as the Fronsdal equation for spin $s=0$
\beq
\boxed{z\pa_z(z\pa_z\varphi)-dz\pa_z\varphi
+\Delta_-\Delta_+\varphi
+ z^2\Box_{(\vx)}\varphi
= 0}
\eeq
up to higher order corrections. As expected, the scalar mass is given by
$$(mL)^2 = -\Delta_-\Delta_+$$ 
Note that the particular value for $\delta$ is picked out by the requirement that the spurious term on the right hand side of equation \eqref{linscalareom2} cancels out. Since $\delta$ was the parameter in the symplectic transformation (\ref{CT1}), we see here the first indication that a symplectic transformation is capable of removing spurious higher order terms, and we will see in section 4.3 that this can be done systematically to all orders.

At the level of the action, we obtain
\beq
S^{(2)}_{bulk} = \int \frac{dz d^d\vx}{z^{d+1}}\Big(\varpi z\pa_z\varphi - d\varpi \varphi -\frac{2(2+2\nu)}{C}\varpi^2-\frac{C}{4}\left(\varphi^2+\alpha z^2\varphi \Box_{(\vx)}\varphi\right)+\cdots\Big)\label{fracps}
\eeq
Solving for the $\varpi$ equation of motion, and plugging it back into the action straightforwardly gives the action (once again up to higher derivative terms)\footnote{We also generate an extra boundary term which can be removed by a boundary counterterm, as a part of holographic renormalization.}  
\beq\label{Fronsdalaction}
\boxed{
S^{(2)}_{bulk} = k\int \frac{dzd^d\vx}{z^{d+1}}\;\Big(z\pa_z\varphi\; z\pa_z\varphi-z^2\varphi\;\Box_{(\vx)}\varphi+(mL)^2\varphi\varphi\Big)+\cdots}
\eeq 
where $k$ is some dimensionful constant.

Having established the bulk action and equations of motion, now let us turn our attention to the boundary terms. Combining equations \eqref{bdryspin0action} and \eqref{bdryspin0action2}, we find that the boundary action is given by
\beq
S_{bdry} = \frac{1}{\epsilon^d}\int d^d\vx\;\varpi(\epsilon,\vx)\left(\varphi(\epsilon,\vx)-\epsilon^{\Delta_-}\phi^{(0)}(\vx)\right)-\frac{\delta}{C\epsilon^d}\int d^d\vx\;\varpi(\epsilon,\vx)\left(1+O(\epsilon^2)\right)\varpi(\epsilon,\vx)
\eeq
This gives rise to the boundary condition
\beq
\varphi -\frac{2\delta}{C}\varpi=\epsilon^{\Delta_-}\phi^{(0)}\left(1+O(\epsilon^2)\right)
\eeq
which upon using the $\varpi$ equation of motion gives
\beq
(z\pa_z-\Delta_+)\varphi
=2\epsilon^{\Delta_-}\phi^{(0)}\left(1+O(\epsilon^2)\right)
\eeq
As usual, as a consequence of the equation of motion \eqref{linscalareom2}, $\varphi$ behaves asymptotically as
\beq
\varphi(z,\vx) = z^{\Delta_+}\varphi^{(+)}(\vx)\left(1+O(z^2)\right)+z^{\Delta_-}\varphi^{(-)}(\vx)\left(1+O(z^2)\right)
\eeq
and the above boundary condition then becomes
\beq
\varphi^{(-)}(\vx) = -\frac{1}{\nu}\phi^{(0)}(\vx)
\eeq
which is the appropriate boundary condition up to a trivial rescaling. For instance in $d=2+1$, we have thus correctly found that the bulk field comes with the ``alternate quantization'' as expected. 

Having warmed up with the spin-zero case, we now generalize the discussion at two levels -- in the next section, we repeat the above exercise for general spin, which will allow us to reproduce the spin-$s$ Fronsdal equation in $AdS_{d+1}$ (once again up to $O(z^4\vpa^4)$ corrections). Then in section \ref{3.3}, we revisit the higher derivative corrections we have been neglecting, and show how to eliminate them systematically. This will complete our argument that the bulk equations obtained from RG are canonically equivalent to $AdS$ Fronsdal equations. 

\subsection{Higher spins}

Moving onto the higher-spin case, we now want to recover the Fronsdal equation for arbitrary spin. As we will show, the computation proceeds in essentially the same way as the $s=0$ case. Going back to the RG equations \eqref{EOMlin1} and \eqref{EOMlin2}, we now wish to turn on bulk fields which are related to the conserved, symmetric and traceless spin-$s$ current in the boundary field theory schematically denoted
$$J^{(s)}_{\mu_1\cdots\mu_s}(\vx) = \;:\phi^*_m\;f_{\mu_1\cdots \mu_s}(\overleftarrow{\pa}, \overrightarrow{\pa})\phi^m:(\vx)$$ 
where $f_{\mu_1\cdots \mu_s}(\vu,\vv)$ is a homogenous, symmetric polynomial of order $s$ in $\vu$ and $\vv$, which is symmetric and traceless in all of its indices. To this end, we choose
\beq
\cscal^{old}_1(z;\vx,\vy) = z^s\phi_{\mu_1\cdots \mu_s}(z,\vx)f^{\mu_1\cdots \mu_s}(\vec{\pa}_{(x)},\vec{\pa}_{(y)})\Big(z^d\delta^d(\vx-\vy)\Big)
\eeq
\beq
\pi^{\mu_1\cdots\mu_s}(z,\vx) = \frac{1}{N}\lim_{\vx\to \vy}\;z^{-s}\;f^{\mu_1\cdots \mu_s}(\vec{\pa}_{(x)},\vec{\pa}_{(y)})\cscalm^{old}_1(z;\vx,\vy) = \frac{1}{N}\langle J^{\mu_1\cdots \mu_s}_{(s)}\rangle_1(z,\vx)
\eeq 
When the current $J_{(s)}^{\mu_1\cdots \mu_s}$ is conserved in the boundary theory, it is clear that the boundary value $\phi_{\mu_1\cdots\mu_s}^{(0)}$ of the source $\phi_{\mu_1\cdots\mu_s}$ is defined only modulo the gauge transformation
\beq
\delta\phi^{(0)}_{\mu_1\cdots \mu_s}(\vx) = \vpa_{(\mu_1}\epsilon^{(0)}_{\mu_2\cdots\mu_s)}(\vx)
\eeq
This is of course a manifestation of the $U(L_2)$ gauge symmetry at the linearized level. Furthermore, since $J_{(s)}$ is traceless, only the traceless part of the boundary source is relevant. We can use these considerations to our advantage by making the gauge choice
\beq
\vpa^{\mu}\phi^{(0)}_{\mu\mu_2\cdots\mu_s}=0, \;\;\;\eta^{\mu_1\mu_2}\phi^{(0)}_{\mu_1\cdots\mu_s}=0\label{bdygaugechoice}
\eeq
For brevity, we introduce the notation $\umu_s \equiv \mu_1\cdots\mu_s$.
The equations of motion \eqref{EOMlin1}, \eqref{EOMlin2} in the present case are given by
\beqn
z\pa_z\phi_{\umu_s}(\vx) &=& \Delta_-\;\phi_{\umu_s}(\vx)\\
z\pa_z\pi^{\umu_s}(\vx) &=& \Delta_+\pi^{\umu_s}(\vx)+\frac{z^{2\nu}}{2}\int d^d\vu\;\dot{G}^{\umu_s,\unu_s}_{(s,s)}(z,\vx,\vu)\phi_{\unu_s}(\vu)\label{spinseompi}
\eeqn
where
\beq
\Delta_+ = d-2+s,\qquad\Delta_- = 2-s,\qquad 2\nu = \Delta_+-\Delta_- = d-4+2s
\eeq
The kernel in eq. (\ref{spinseompi}) can be identified with
\beqn
G_{(s,s)}^{\umu_s,\unu_s}(z;\vx,\vy) &=& \frac{2i}{N}\left\langle J_{(s)}^{\mu_1\cdots\mu_s}(\vx)J_{(s)}^{\nu_1\cdots\nu_s}(\vy)\right\rangle_{CFT,Mink}(z)\\
\dot{G}_{(s,s)}^{\umu_s,\unu_s}(z;\vx,\vy) &=& z\pa_zG_{(s,s)}^{\umu_s,\unu_s}(z;\vx,\vy)
\eeqn 
where the correlator is defined in the regulated CFT on Minkowski space. To avoid cluttering the notation, we will drop the subscript $(s,s)$ on these kernels henceforth.

Remarkably, the equations of motion are compatible with the gauge choice on the boundary, which implies that we can take the bulk fields (or more precisely, on-shell bulk fields) to satisfy the same gauge conditions 
\beq
\vpa^{\mu}\phi_{\mu\mu_2\cdots\mu_s}=0=\vpa_{\mu}\pi^{\mu\mu_2\cdots\mu_s},\; \;\;\;\eta^{\mu_1\mu_2}\phi_{\mu_1\cdots\mu_s}=0=\eta_{\mu_1\mu_2}\pi^{\mu_1\cdots\mu_s}\label{bulkgaugechoice}
\eeq
This choice of (on-shell) gauge is once again the higher-spin \emph{Coulomb gauge} (see Appendix A) at the level of RG. The above equations of motion come from the action 
\beq
S^{(2)}_{bulk} = \int \frac{dz d^d\vx}{z^{d+1}}\;\left(\pi^{\umu_s}(z,\vx) z\pa_z\phi_{\umu_s}(z,\vx)-\Delta_-\pi^{\umu_s}(z,\vx) \phi_{\umu_s}(z,\vx) +\frac{z^{2\nu}}{4}\phi_{\umu_s}(z,\vx)\dot{G}^{\umu_s,\unu_s}(z;\vx,\vy) \phi_{\unu_s}(z,\vy)\right)
\eeq
along with the boundary action
\beq
S_{bdry} = \frac{1}{\epsilon^d}\int d^d\vx\;\pi^{\umu_s}(\epsilon,\vx)\left(\phi_{\umu_s}(\epsilon,\vx)-\epsilon^{\Delta_-}\phi^{(0)}_{\umu_s}(\vx)\right)
\eeq

Let us pause briefly to explain why the higher-spin Coulomb gauge simplifies the analysis significantly. As before, the kernel $\dot{G}^{\umu_s,\unu_s}$ admits an asymptotic expansion, which in general is complicated because of the index structure. But precisely in this gauge (\ref{bulkgaugechoice}), we see from the action above that the index structures become irrelevant; the only part of the kernels which survive in the action take the generic form
\beqn\label{HSG1}
\hsdG^{\umu_s,\unu_s}(\vx,\vy)&=&-C_sz^{-2\nu}\left(1+\alpha_s z^2\Box_{(\vx)}+\cdots\right)\eta^{\langle \mu_1\langle \nu_1}\cdots\eta^{\mu_s\rangle \nu_s\rangle }\;\delta^d(\vx-\vy)
\\
\label{HSG2}
\hsdG^{-1}_{\umu_s,\unu_s}(\vx,\vy) &=& -\frac{z^{2\nu}}{C_s}\left(1-\alpha_s z^2\Box_{(\vx)}+\cdots\right)\eta_{\langle \mu_1\langle \nu_1}\cdots\eta_{\mu_s\rangle \nu_s\rangle }\;\delta^d(\vx-\vy)
\eeqn
where $\alpha_s> 0$ and $C_s$ are (dimensionful) constants (see Appendix B).\footnote{While in the present discussion $\alpha_s>0$ is required for the bulk metric to have the correct signature, one could imagine having a cut-off function where this condition is not satisfied. The more general argument of section 4.3 will show that this condition (namely $\alpha_s>0$) is not actually necessary -- it is merely an artifact of the simple-minded canonical transformation we have chosen here.} The notation $\langle \mu_1\cdots \mu_s\rangle$ denotes the symmetrized traceless combination, and the ellipsis above indicate higher-derivative terms.

Moving on, we now perform the canonical transformation
\beqn
\phi_{\umu_s}(z,\vx)&=& \varphi_{\umu_s}(z,\vx)+ \frac{2\delta }{z^{2\nu}}\int_{\vy}\hsdG^{-1}_{\umu_s,\unu_s}(z;\vx,\vy)\varpi^{\unu_s}(z,\vy)\nonumber\\
\pi^{\umu_s}(z,\vx) &=& \varpi^{\umu_s}(z,\vx)
\eeqn
for some constant $\delta$ to be fixed later. This canonical transformation preserves the higher-spin Coulomb gauge condition
$
\vpa^{\mu}\varphi_{\mu\mu_2\cdots\mu_s}=0, \;\eta^{\mu_1\mu_2}\varphi_{\mu_1\cdots\mu_s}=0
$, 
as can be easily checked.

In terms of the new fields, the action becomes
\beqn
S^{(2)}_{bulk} &=& \int \frac{dz}{z^{d+1}}\;\Big(\varpi^{\umu_s} \bl z\pa_z\varphi_{\umu_s} -(\Delta_--\delta)\varpi^{\umu_s}\bl\varphi_{\umu_s}+ \frac{1}{z^{2\nu}}\varpi^{\umu_s}\bl\left[\delta^2\hsdG_{\umu_s,\unu_s}^{-1}+\delta z\pa_z\hsdG^{-1}_{\umu_s,\unu_s}\right]\bl\varpi^{\unu_s}\nonumber\\
&&+\frac{1}{4z^{-2\nu}}\varphi_{\umu_s}\bl \hsdG^{\umu_s,\unu_s}\bl \varphi_{\unu_s}\Big)
\\
S_{bdry} &=& \frac{1}{\epsilon^{d}}\int d^d\vx\;\varpi^{\umu_s}(x)\left(\varphi_{\umu_s}(\vx)-\epsilon^{\Delta_-}\phi^{(0)}_{\umu_s}(\vx)\right)-\frac{\delta}{\epsilon^{d+2\nu}}\left.\varpi^{\umu_s}\bl \hsdG^{-1}_{\umu_s,\unu_s}\bl \varpi^{\unu_s}\right|_{z=\epsilon}
\eeqn
Substituting equations \eqref{HSG1} and \eqref{HSG2} into the above action, we find that the $\varpi^2$ term in the action becomes 
\beq
\frac{1}{C_s}\int \frac{dzd^d\vx}{z^{d+1}}\varpi^{\umu_s}(z,\vx)\Big(-\delta(\delta+2\nu)+\alpha_s \delta(\delta+2\nu+2) z^2\Box_{(\vx)}+\cdots\Big)\varpi_{\umu_s}(z,\vx)
\eeq
As in the $s=0$ case above, choosing $\delta = -(2\nu+2)$ will ensure that the $\varpi\Box_{(\vx)}\varpi$ term drops out, and the full bulk action then becomes
\beqn
S^{(2)}_{bulk} &=& \int \frac{dzd^d\vx}{z^{d+1}}\;\Big(\varpi^{\umu_s} z\pa_z\varphi_{\umu_s} -(d+s)\;\varpi^{\umu_s}\varphi_{\umu_s}- \frac{1}{C_s}2(\Delta_++s)\;\varpi^{\umu_s} \varpi_{\umu_s}\nonumber\\
&-&\frac{C_s}{4}\varphi_{\umu_s}\left(1+\alpha_sz^2\Box_{(\vx)}\right)\varphi^{\umu_s}\Big)+\cdots\label{HSHam}
\eeqn
The equations of motion for this action are now
\beqn\label{HSEOM1}
z\pa_z\varphi_{\umu_s}-(d+s)\;\varphi_{\umu_s} &=& \frac{2}{C_s}2(\Delta_++s)\varpi_{\umu_s}+\cdots
\\
\label{HSEOM2}
-z\pa_z\varpi^{\umu_s}-s\varpi^{\umu_s} &=& \frac{C_s}{2}\left(1+\alpha_s\frac{z^2}{M^2}\Box_{(\vx)}\right)\varphi^{\umu_s}+\cdots
\eeqn
Combining these two equations into a second order differential equation, we get (up to higher derivative terms)
\beq
z\pa_z\left(z\pa_z\varphi_{\umu_s}\right)- d\;z\pa_z\varphi_{\umu_s} +\ell_s^2z^2\Box_{(\vx)}\varphi_{\umu_s}-s(s+d)\varphi_{\umu_s}+2(\Delta_++s)\varphi_{\umu_s}=0 \label{FronsdalEOM'}
\eeq
where $\ell_s^2 = \frac{2(\Delta_++s)\alpha_s}{M^{2}}$ is a positive constant. As before, $\ell_s$ can be set equal to one, by rescaling the boundary coordinate $\vx$. Finally, in order to put the above equation in the standard Fronsdal form, we redefine
\beq
\varphi_{\umu_s} = z^s\widehat{\varphi}_{\umu_s}
\eeq
We note that this is not an arbitrary redefinition, but corresponds to going from frame indices to coordinate indices. Having done so, the above equation in terms of $\widehat{\varphi}_{\umu_s}$ becomes
\beq
\boxed{z\pa_z\left(z\pa_z\widehat{\varphi}_{\umu_s}\right)+(2s-d)\;z\pa_z\widehat{\varphi}_{\umu_s} +z^2\Box_{(\vx)}\widehat{\varphi}_{\umu_s}+\left[s(s-d)+\Delta_+\Delta_-\right]\widehat{\varphi}_{\umu_s}=0}\label{FronsdalEOM}
\eeq
which is precisely the Fronsdal equation in the higher-spin Coulomb gauge (see equation \eqref{Coulombeom}). It is worth pointing out that in the special case $s=1$ this is the familiar Maxwell's equation in $AdS$ space written in Coulomb gauge, and the Hamiltonian obtained from equation \eqref{HSHam} can be cast in the form $\vec{E}^2+\vec{B}^2$. Similarly, in the case $s=2$ the above equation is the Einstein's equation linearized about $AdS$ space, in the $s=2$ Coulomb gauge.

Finally, we revisit the boundary action
\beq
S_{bdry} = \frac{1}{\epsilon^{d}}\int d^d\vx\;\varpi^{\umu_s}(\epsilon,\vx)\left(\varphi_{\umu_s}(\epsilon,\vx)-\epsilon^{\Delta_-}\phi^{(0)}_{\umu_s}(\vx)\right)-\frac{\delta}{C_s\epsilon^{d}}\int d^d\vx\;\varpi^{\umu_s}(\epsilon,\vx) \left(1+O(\epsilon^2)\right) \varpi_{\umu_s}(\epsilon,\vx)
\eeq
which gives us the boundary condition 
\beq
\varphi_{\umu_s} - \frac{2\delta}{C_s}\varpi_{\umu_s} = \epsilon^{\Delta_-}\phi^{(0)}_{\umu_s}\left(1+O(\epsilon^2)\right)
\eeq
Using $\delta=-2\nu-2=-(\Delta_++s)$ and the equation of motion \eqref{HSEOM1}, we get
\beq
z\pa_z\varphi_{\umu_s}-\Delta_+\varphi_{\umu_s}=2\epsilon^{\Delta_-}\phi_{\umu_s}^{(0)}\left(1+O(\epsilon^2)\right)
\eeq
Equation \eqref{FronsdalEOM'} implies the asymptotics 
$$\lim_{z\to 0}\varphi_{\umu_s}(z,\vx) \sim \varphi^{(+)}_{\umu_s}(\vx)z^{\Delta_+}\left(1+O(z^2)\right)+\varphi_{\umu_s}^{(-)}(\vx)z^{\Delta_-}\left(1+O(z^2)\right)$$
Therefore, the boundary condition becomes
\beq
\varphi^{(-)}_{\umu_s} = -\frac{1}{\nu}\phi^{(0)}_{\umu_s}
\eeq
or equivalently $\widehat{\varphi}_{\umu_s}\sim -\frac{z^{2-2s}}{\nu}\phi^{(0)}_{\umu_s}$, which is indeed the correct boundary condition up to a trivial rescaling. 
 
\subsection{Higher order terms}\label{3.3}
So far we have demonstrated that the linearized bulk equations obtained from RG are canonically equivalent to $AdS_{d+1}$ Fronsdal equations, up to $O(z^4\vpa^4)$ terms. These higher derivative terms are only an artifact of choosing a simple canonical transformation. Indeed, it is possible to construct a more general canonical transformation such that the higher derivative terms are completely eliminated, as we will now show. For notational simplicity, we revert back to the spin zero case; all the arguments carry through straightforwardly in the general spin case. So consider once again a general linear canonical transformation 
\beq
\phi = A\bl\varphi + B\bl\varpi
\eeq
\beq
\pi = C\bl\varphi+D\bl\varpi
\eeq
where we take all the matrices $A,B,C,D$ to be symmetric as well as translationally and rotationally invariant. The requirement that this be a canonical transformation gives us one constraint
\beq\label{c1}
A\bl D - C\bl B = \boldsymbol{1}
\eeq
where $\boldsymbol{1}$ of course is the delta function $\delta^d(\vx-\vy)$. The original bulk action \eqref{spin0action} in terms of the new variables is given by
\beqn
S^{(2)}_{bulk} &=& \int \frac{dz}{z^{d+1}}\Big\{\varpi\bl z\pa_z\varphi-\varpi\bl\left((\dot{C}-\Delta_+C)\bl B-D\bl (\dot{A}-\Delta_-A)-\frac{z^{2\nu}}{2}B\bl \dot{G}\bl A\right)\bl\varphi\nonumber\\
&-&\frac{1}{2}\varphi\bl\left(\dot{C}\bl A-C\bl \dot{A}-2\nu(C\bl A)-\frac{z^{2\nu}}{2}A\bl\dot{G}\bl A\right)\bl \varphi\\
&-&\frac{1}{2}\varpi\bl\left(\dot{D}\bl B - D\bl \dot{B}-2\nu(D\bl B)-\frac{z^{2\nu}}{2}B\bl\dot{G}\bl B\right)\bl\varpi\Big\}\nonumber
\eeqn
with additional boundary terms coming from the integrations by parts we have performed above
\beq
\delta S_{bdry} = \frac{1}{2\epsilon^d}\Big(\varphi\bl(C\bl A)\bl \varphi+\varpi\bl (D\bl B)\bl \varpi+2\varphi\bl (C\bl B)\varpi\Big)
\eeq
Here $\dot{A} = z\pa_zA$, and recall the definitions
$$ \Delta_+ = d-2,\;\;\Delta_- = 2, \;\; 2\nu = \Delta_+-\Delta_-$$
relevant to $s=0$. Remember that our aim here is to map this action on to the Klein-Gordon action in \eqref{fracps}, with no higher-derivative corrections surviving. So this gives us three more constraints:
\beq\label{c2}
(\dot{C}-\Delta_+C)\bl B-D\bl (\dot{A}-\Delta_-A)-\frac{z^{2\nu}}{2}B\bl \dot{G}\bl A=d\;\boldsymbol{1}
\eeq
\beq\label{c3}
\dot{C}\bl A-C\bl \dot{A}-2\nu(C\bl A)-\frac{z^{2\nu}}{2}A\bl\dot{G}\bl A = \frac{C_0}{2}\left(1+\alpha z^2\Box_{(\vx)}\right)\boldsymbol{1}
\eeq
\beq\label{c4}
\dot{D}\bl B - D\bl \dot{B}-2\nu(D\bl B)-\frac{z^{2\nu}}{2}B\bl\dot{G}\bl B = \frac{4(2+2\nu)}{C_0}\boldsymbol{1}
\eeq
where $\alpha$ and $C_0$ are constants. Together with the symplectic constraint $A\bl D-C\bl B = \boldsymbol{1} $, we now have four constraints and four unknown kernels --- so we can try to solve for them order by order in an asymptotic expansion in powers of $z^2\Box_{(\vx)}$. Of course, we have already found the solution to these constraints up to second order in derivatives previously, so we might as well retain the previous solution up to two derivatives. We parametrize the higher derivatives as follows:
\beq\label{e1}
A =\delta^d(\vx-\vy)+ \Big(\alpha^A_2\; z^4\Box_{(\vx)}^2+\alpha^A_3\;z^6\Box_{(\vx)}^3+\cdots\Big)\delta^d(\vx-\vy)
\eeq
\beq\label{e2}
B = -2(2+2\nu)\left(\frac{1}{\alpha^G_0}-\frac{\alpha_1^G}{(\alpha^G_0)^2}z^2\Box_{(\vx)}\right)\delta^d(\vx-\vy)+\Big(\alpha_2^B\;z^4\Box_{(\vx)}^2+\alpha^B_3\;z^6\Box_{(\vx)}^3+\cdots\Big)\delta^d(\vx-\vy)
\eeq
\beq\label{e3}
C = \Big(\alpha^C_2\;z^4\Box^2_{(\vx)}+\alpha^C_3\;z^6\Box_{(\vx)}^3+\cdots \Big)\delta^d(\vx-\vy)
\eeq
\beq\label{e4}
D = \delta^d(\vx-\vy)+\Big(\alpha^D_2\;z^4\Box_{(\vx)}^2+\alpha^D_3\;z^6\Box_{(\vx)}^3+\cdots\Big)\delta^d(\vx-\vy)
\eeq
where $\boldsymbol{\alpha}^{(i)} = (\alpha^A_i, \alpha^B_i, \alpha^C_i, \alpha^D_i)$ for $i \geq 2$ are coefficients to be determined from the constraints. We have also introduced the convenient notation
\beq
\dot{G}_{(0,0)}(z;\vx,\vy) = z^{-2\nu}\Big(\alpha^G_0+\alpha^G_1\;z^2\Box_{(\vx)}+\cdots\Big)\delta^d(\vx-\vy) \label{dotg}
\eeq
with $\alpha^G_0\neq 0$. Note that we have taken the expansions for $A, B, C, D$ to be polynomial in $z^2\Box_{(\vx)}$. While this is correct in odd dimensions, in general one needs to include logarithmic terms in even dimensions. In order to avoid such complications, we have restricted our attention to odd dimensions in this paper; the same arguments should go through in even dimensions with logarithmic terms properly taken into account. 

The game now is to determine the coefficients $\boldsymbol{\alpha}^{(i)}$. Let us describe this process in general. Let's say we have determined the coefficients to the $(r-1)$th order in the above expansion. At the $rth$ order ($r \geq 2$), we now have four variables $\alpha^A_r,\cdots \alpha^D_r$ to determine, from the four constraints (\ref{c1}, \ref{c2}--\ref{c4}) listed above. Plugging our expansions (\ref{e1}--\ref{dotg}) into the constraints, we get four constraint equations on the coefficients $\boldsymbol{\alpha}^{(r)} = (\alpha^A_r, \alpha^B_r, \alpha^C_r, \alpha^D_r)$:

1. Symplectic contraint:
\beq
\alpha^A_r+\alpha^D_r+\frac{2(2+2\nu)}{\alpha^G_0}\alpha^C_r = f^{(r)}_1
\eeq

2. $\varpi\varphi$ constraint:
\beq
\frac{2(2+2\nu)}{\alpha^G_0}(2r-\Delta_+)\alpha^C_r+(2r-2-\Delta_+)\alpha^A_r-\Delta_-\alpha^D_r+\frac{\alpha^G_0}{2}\alpha^B_r = f^{(r)}_2
\eeq

3. $\varphi^2$ constraint:
\beq
(2r-2\nu)\alpha^C_r - \alpha^G_0\alpha^A_r = f^{(r)}_3
\eeq

4. $\varpi^2$ constraint:
\beq
-\frac{2(2+2\nu)}{\alpha^G_0}(2r-2\nu)\alpha^D_r-(2r-2\nu-4)\alpha^B_r = f^{(r)}_4
\eeq
where on the right hand side we have functions $\boldsymbol{f}^{(r)} = (f^{(r)}_1, \cdots, f^{(r)}_4)$ of all the previously determined coefficients and $\{\alpha^G_j\}$, i.e., $\boldsymbol{f}^{(r)} = \boldsymbol{f}^{(r)}(\boldsymbol{\alpha}^{(0)},\cdots, \boldsymbol{\alpha}^{(r-1)};\{\alpha^G_j\})$. So the general structure of these equations for any $r$ is given by 
\beq\label{deteq}
\boldsymbol{M}^{(r)}\cdot \boldsymbol{\alpha}^{(r)} = \boldsymbol{f}^{(r)}
\eeq 
where
\beq
\boldsymbol{M}^{(r)}= \left(\begin{matrix}
1 & 0 & \frac{2(2+2\nu)}{\alpha^G_0} & 1\\
(2r-2-\Delta_+) & \frac{\alpha^G_0}{2} & \frac{2(2+2\nu)}{\alpha^G_0}(2r-\Delta_+) & -\Delta_-\\
-\alpha^G_0 & 0 & (2r-2\nu) & 0\\
0 & -(2r-2\nu-4) & 0 & -\frac{2(2+2\nu)}{\alpha^G_0}(2r-2\nu)
\end{matrix}
\right)
\eeq
and $\boldsymbol{\alpha}^{(r)} = (\alpha^A_r,\alpha^B_r,\alpha^C_r, \alpha^D_r)$, $\boldsymbol{f}^{(r)} = (f^{(r)}_1,\cdots, f^{(r)}_4)$ are defined above. The above matrix has the determinant
\beq
\mathrm{det}\;\boldsymbol{M}^{(r)}  = -8r(r-\nu)(r+\nu)
\eeq
We see that the determinant is non-zero for generic $r>0$, except at the pathological levels $r = |\nu|$, where the determinant vanishes. However, $r$ is an integer, while for $d$ odd, $\nu$ is always half-integral -- hence there are no pathologies for any $r>0$ when $d$ is odd. Consequently, $\mathrm{det}\;\boldsymbol{M}^{(r)} \neq 0$ for any $r>0$, which means that we can solve equation \eqref{deteq} to obtain $\boldsymbol{\alpha}^{(r)}$. By induction on $r$, we can thus determine all the coefficients of the kernels $A, B, C, D$ uniquely, and determine the canonical transformation at any desired order in the asymptotic expansion. While we demonstrated this in the case of $s=0$ above, the same calculation generalizes straightforwardly for general spin with the same conclusion. This completes our proof of the statement that in all odd dimensions, the RG equations are canonically equivalent to the bulk Fronsdal equations. 

A few comments are in order: firstly, if we naively carry over all the above expressions to $d$ even, then it might seem that the program fails at $r = |\nu|$. This indicates that the asymptotic form of the expansions for $A,B,C,D$ we have considered above is incomplete for $d$ even -- we must also include terms logarithmic in $z^2\vpa^2$. Having done so, the arguments we have presented above will go through for even dimensions as well, but we will not repeat the details here. Secondly, our discussion does not crucially depend on the choice of the cut-off function $K_F$ -- as long as $\dot{G}$ has an expansion of the form \eqref{dotg}, all the arguments go through. Of course, the detailed form of the canonical transformation would depend on the choice of the cut-off function. From this point of view, we conclude that the various different choices of cut-off functions in the boundary correspond to different choices of a canonical-frame in the bulk. Finally, we note that although we have shown the existence of the canonical transformation to all orders in the expansion in powers of $z^2\vpa^2$, these expansions are still somewhat formal, i.e., we do not have any handle on the convergence of the series we have found for $A, B,C, D$.

\section{Discussion}\label{section5}
In conclusion, we have shown that the linearized exact renormalization group equations for free $U(N)$ vector models in the single-trace sector are precisely canonically equivalent to the higher-spin Fronsdal equation in $AdS$ space. We will end with some speculative comments and open problems:

(i) \textbf{Bulk Locality}: Although it is widely believed that $AdS$/CFT provides a modern, geometric viewpoint on the renormalization group, it is somewhat puzzling how non-local\footnote{We say non-local because turning on a certain single-trace, quasiprimary operator generates infinitely many other operators along the flow, which are important to keep track of in the context of holography. We were able to do this systematically in the exact RG formlism for free vector models, without any discrimination between relevant, marginal and irrelevant operators.}  RG equations can be equivalent to local bulk equations. We have seen above in the case of the free vector model/higher-spin duality, that there exists a canonical frame in which the linearized RG equations give rise to local, second order differential equations in the bulk for individual local spin-$s$ fields, namely the Fronsdal equations. A satisfying feature is that this result is not really sensitive to the detailed form of the chosen cut-off function. The canonical transformation can be thought of as giving us the correct renormalization scheme in which the bulk is local (at the linearized level). It will be interesting to try and extend these ideas to more general interacting CFTs, where the question of bulk locality becomes more significant \cite{Heemskerk:2009pn}. 

(ii) \textbf{Bulk interactions}: Another interesting question is whether these results can be extended beyond the linearized level -- more precisely, can we match the cubic interactions obtained from RG with the cubic interactions in Vasiliev theory? Thinking along the lines of \cite{Gopakumar:2003ns} might be fruitful in this case. Of course, we do not expect the bulk to be local beyond cubic order, and ``deriving'' the Vasiliev theory from RG is an open problem.

(iii) \textbf{Gauge interactions \&\ String field theory}: So far we have only focussed on the case of free vector models with global $U(N)$ symmetry. An extremely interesting possibility is to turn on gauge fields in the boundary CFT, such as for example in the $d=2+1$ Chern-Simons-vector models \cite{Giombi:2011kc,Chang:2012kt}. As we noted in section 2, one must be more careful in defining gauge-invariant bilocal operators in the presence of gauge interactions. The bilocal operator $\hat{\Pi}(\vx,\vy)=\phi_m^*(\vx)\phi^m(\vy)$ in the ordinary vector model must now be improved by the inclusion of a $U(N)$ Wilson line between the two vectors, and thus becomes a functional of open strings:
\beq
\hat{\Pi}[\vx^{\mu}(\sigma)] =   \phi_m^*(\vx^{\mu}(0))\mathscr{P}{\exp\left(\int_{0}^{\pi}d\sigma' \;A_{\mu}[\vx(\sigma')]\dot{\vx}^{\mu}(\sigma')\right)^m}_n\phi^n(\vx^{\mu}(\pi))
\eeq
where $A_{\mu}$ is the $U(N)$ gauge field. Correspondingly, the bi-local source $B(\vx,\vy)$ in the free vector model now becomes an open-string functional in the Chern-Simons vector model:
\beq
B[\vx^{\mu}(\sigma)]
\eeq
For large but finite Chern-Simons level $k$, it is then natural to conjecture that the exact renormalization group equations for $B[\vx(\sigma)]$ and $\Pi[\vx(\sigma)]$ should be interpreted as open-string field equations in $AdS$ space, in a Hamiltonian form (see the discussion in footnote \ref{ftnt} for further inspiration).\footnote{More precisely, one must also include closed Wilson-loop operators and corresponding sources in the boundary, and expect to have an open-closed string theory in the bulk.} Pictorially, this corresponds to filling in the Witten diagrams of the vector model (fig 1) to obtain open-string worldsheets in a gauge where the worldsheet time is identified with the radial coordinate of $AdS$. Naturally, if we take the Chern-Simons level $k\to \infty$, then the Chern-Simons action localizes on flat connections, and the dependence on the strings drops out. In this limit $B$ and $\Pi$ only depend on the end-points, and we thus collapse back to the bi-local sources and operators of the ordinary $U(N)$ vector model. In this sense, the bulk theory dual to free-vector models we described in section 2 (in terms of bi-locals) should be thought of as a certain tensionless limit of open-string field theory in $AdS$. (See also \cite{Lee:2011wx} for related discussion, and \cite{Gopakumar:2004qb,Gopakumar:2005fx} for a different approach to emergence of $AdS$ strings from free gauge theory.)  

(iv) \textbf{Entanglement renormalization}: There has also been an interesting proposal \cite{PhysRevD.86.065007, Nozaki:2012zj} that the tensor network construction of ground states of critical systems -- MERA (Multi-Scale Entanglement Renormalization) -- is closely related to holography. The idea is that coarse-graining a state by progressively removing entanglement at longer and longer length scales (i.e. entanglement renormalization) via the action of unitary operations (``disentanglers'') gives rise to a holographic description of the critical system. It would be very interesting to connect the ideas in this paper with the idea of entanglement renormalization; more precisely, are the canonical transformations we described within the conventional RG language related to the disentanglers in the MERA description? We leave these questions for future work.    

\section*{Acknowledgements}
We are grateful to Cheng Peng for early collaboration, and to Sung-Sik Lee, Eric Mintun, Tassos Petkou and Joe Polchinski for helpful discussions, and Xavier Bekaert for an email clarification. OP would like to thank the Kavli Institute for Theoretical Physics (Santa Barbara) for their hospitality, where a part of this work was completed. This research was supported in part by
the U.S. Department of Energy contract DE-FG02-13ER42001. In addition, OP was supported in part by a graduate fellowship at KITP, under the U.S. National Science Foundation grant number NSF PHY11-25915.

\appendix

\section{The Higher Spin Coulomb gauge}
\label{app:gauge}
In this appendix, we discuss the higher spin Coulomb gauge for the linearized higher-spin fields $\vh_{I_1\cdots I_n}$, i.e.
\beq
\vh_{z\cdots z\mu_1\cdots \mu_s} = 0\;\;\cdots (s < n),\;\;\; \pa_{\mu}{\vh^{\mu}}_{\mu_2\cdots \mu_{s-1}}=0,\;\;{\vh^{\mu}}_{\mu\mu_3\cdots \mu_s}=0
\eeq
The Fronsdal equation (without any gauge fixing) is given by
\beq
\nabla_I \nabla^I \vh_{I_1...I_n} - n \nabla_I \nabla_{I_1} \vh^I {}_{I_2 ... I_n}
+\tfrac{1}{2}n(n-1) \nabla_{I_1} \nabla_{I_2} \vh^I {}_{I I_3...I_n}-2(n-1)(n+d-2) \vh_{I_1 ... I_n} = 0 
\label{fronsdala}
\eeq
where the fields $\vh_{I_1\cdots I_n}$ are double-traceless. We begin by considering the \emph{de Donder}  condition
\begin{equation}
(\nabla \cdot \vh)_{I_2 ... I_n} - \frac{n-1}{2} \nabla_{(I_2} \vh' {}_{I_3 ... I_n)} = 0 \ ,
\label{deDonder}
\end{equation}
where $\vh'_{I_3\cdots I_n} = {\vh^I}_{II_3\cdots I_n}$. If the higher spin fields were massive, this would follow from the equations of motion. In the `massless' case which we are concerned with here, the de Donder condition can be chosen as a gauge condition. The equations of motion \eqref{fronsdala} in this gauge simplify to
\beq
\nabla^I \nabla_I \vh_{I_1...I_n} 
 -\Big(2(n-1)(d-2+n)-  n(d-1+n)\Big) \vh_{I_1 ... I_n}
 -n(n-1) g_{(I_1I_2}  \vh'_{I_3 ... I_n)}
 = 0
\label{fronsdal1}
\eeq
The gauge transformations which preserve this gauge satisfy
\begin{equation}
\left[ \nabla_I \nabla^I - (n-1)(n+d-2) \right] \Lambda_{I_2 ... I_n} = 0.
\label{constraint}
\end{equation} 
Further, it was shown in \cite{Mikhailov:2002bp}, that for \emph{on-shell} Fronsdal fields the trace ${\vh^I}_{II_3\cdots I_n}$ can be gauged away, and we thus arrive at the on-shell de Donder condition
\begin{equation}
(\nabla \cdot \varphi)_{I_2 ... I_n} = 0 \ , \qquad  \varphi' {}_{I_3 ... I_n} = 0.
\label{sdeDonder}
\end{equation}
In this gauge, the Fronsdal equation simplifies greatly
\begin{equation}
\Big(\nabla_I \nabla^I - (n^2 + (d-5)n-2(d-2)) \Big) \varphi_{I_1 ... I_n} = 0 \ .
\label{fronsdal2}
\end{equation}
It is worth pointing out that the on-shell de Donder gauge is the analog of the Lorentz gauge in the case of spin one fields. 

However, we have still not isolated the physical degrees of freedom. To see why, it is helpful to think about the spin one case. Naively, in $D$ spacetime dimensions, the spin-one gauge field $A_{I}$ has $D$ degrees of freedom. The Lorentz gauge condition $\nabla_IA^I=0$ reduces this to $(D-1)$ -- however, the number of physical degrees of freedom carried by a spin-one gauge field is actually $(D-2)$. This is because closer inspection reveals that one of the components of the gauge field (which in the context of $AdS$/CFT, we may take to be the $z$ component $A_z$) does not contribute to the symplectic structure. In simpler terms, $A_z$ does not have a canonical momentum, and is thus not a dynamical field, but a Lagrange multiplier which enforces the Gauss' law constraint. As long as we satisfy the equation of motion for $A_z$ (namely the Gauss' law), we are at the liberty to set $A_z =0$. We thus arrive at the physical degrees of freedom carried by the remaining components $A_{\mu}$, which further satisfy the Coulomb gauge condition $\vpa_{\mu} A^{\mu}=0$. Of course, in the $AdS$/CFT context, this is precisely the right number of degrees of freedom for a conserved spin-one current in the boundary CFT!

The same story generalizes to the case of higher-spin fields. Closer inspection of the on-shell de Donder gauge condition reveals that the conjugate momenta of all the higher-spin gauge fields of the form $\vh_{z\cdots z\mu_1\cdots \mu_s}$ for $s<n$ can be written in terms of the spatial divergence of other fields:
\beq
\nabla_z{\vh^z}_{I_2\cdots I_n} = -\nabla_{\mu}{\vh^{\mu}}_{I_2\cdots I_n}
\eeq
Therefore, these fields do not contribute to the symplectic structure, and are non-dynamical. We then have the freedom to set $\vh_{z\cdots z\mu_1\cdots \mu_s}=0$ for $s<n$. Happily, the corresponding equations of motion are straightforwardly satisfied upon making this choice, and therefore this choice is a consistent truncation of the Fronsdal equations. Indeed, the remaining physical fields, namely $\vh_{\mu_1\cdots \mu_s}$, which satisfy
\beq
\vpa_{\mu}{\vh^{\mu}}_{\mu_2\cdots \mu_s} = {\vh^{\mu}}_{\mu\mu_3\cdots \mu_s}= 0
\eeq
carry precisely the right number of degrees of freedom as the conserved, spin-$s$ quasi-primary operator in the dual CFT. The Fronsdal equation written in terms of these fields becomes
\beq
z^2\pa_z^2\varphi_{\mu_1\cdots\mu_s}+(2s-d+1)z\pa_z\varphi_{\mu_1\cdots\mu_s}+z^2\Box_{(\vx)}\varphi_{\mu_1\cdots\mu_s}+\left[s(s-d)+(d-2+s)(2-s)\right]\varphi_{\mu_1\cdots\mu_s}=0
\eeq

\section{Calculating $\dot{G}_{(s,s)}$}\label{appB}
In this appendix, we want to explicitly compute the kernel $\dot{G}_{(s,s)}$ which appeared in section 4, equations \eqref{G}, \eqref{HSG1}. We will first compute the $s=0$ case, and then general $s$. 
\subsection{$s=0$}
From the definition \eqref{defG}, we have
\beq
\udG(z;\vx,\vy) = \int \frac{d^d\vec{p}}{(2\pi)^d}\;\udG(z;\vp)e^{i\vp.(\vx-\vy)},\;\;\;\udG(z;\vp) = c\int\frac{d^d\vq}{(2\pi)^d}\frac{K(z^2(\vp-\vq)^2/M^2)}{(\vp-\vq)^2}\frac{K(z^2\vq^2/M^2)}{\vq^2}
\eeq
where $c$ is some constant factor. This is basically the Feynman diagram shown in figure 2.
\begin{figure}[!h]
\centering
\includegraphics[height=3.5cm, width=6cm]{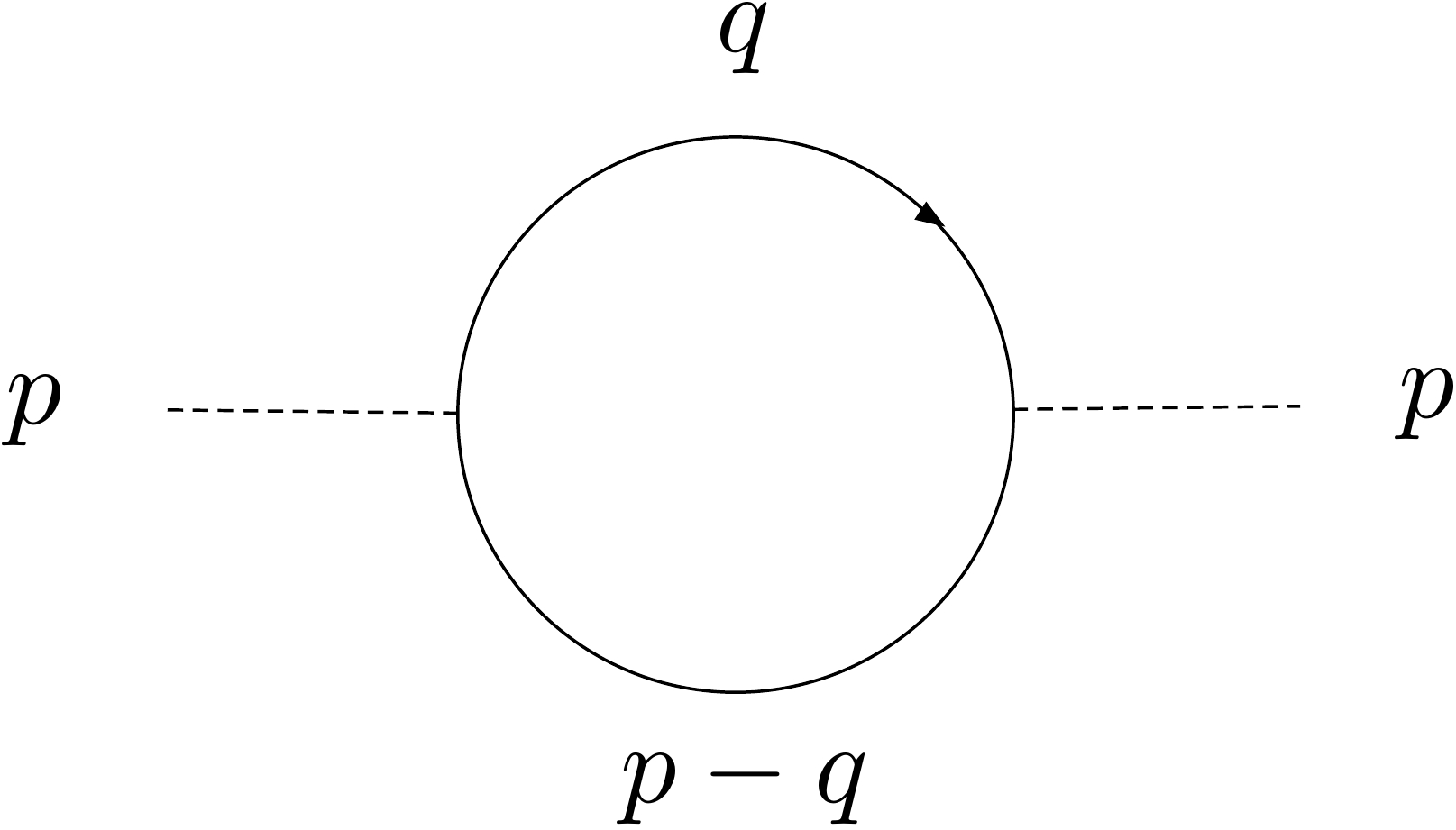}
\caption{\small{The Feynman diagram which enters the renormalization group equations at the linearized level. The dotted lines are the external sources, while the solid lines correspond to propagators for elementary scalars.}}
\end{figure}

For concreteness, let us pick a convenient regulator:
$$K(s) = e^{-s}$$
As we have discussed before, the arguments we have presented do not depend on the choice of the cut-off function. Therefore
\beq
\udG(z;\vp) = c\int\frac{d^d\vq}{(2\pi)^d}\frac{e^{-u^2(\vp-\vq)^2}}{(\vp-\vq)^2}\frac{e^{-u^2\vq^2}}{\vq^2}
\eeq
where we have defined 
$$u=z/M$$
We can use Schwinger parameters to rewrite this as
\beq
\udG(z;\vp) = c\int_{u^2}^{\infty}\int_{u^2}^{\infty}dtds\int\frac{d^d\vq}{(2\pi)^d}e^{-t(\vp-\vq)^2-s\vq^2}=c\int_{u^2}^{\infty}\int_{u^2}^{\infty}dtds\;\frac{1}{2^d\pi^{d/2}}\frac{1}{(s+t)^{d/2}}e^{-\frac{ts}{t+s}\vp^2}
\eeq
where we have carried out the $\vq$ integration. We can evaluate the $u=0$ limit straightforwardly
\beq
\udG(z\to0;\vp) = c\frac{\Gamma(2-\frac{d}{2})B(\frac{d}{2}-1,\frac{d}{2}-1)}{(4\pi)^{d/2}}\frac{1}{(\vp^2)^{2-\frac{d}{2}}}
\eeq
which in position space goes as $|x-y|^{-2\Delta_+}$ -- the correct boundary two point function. But what we are interested in is not $\udG$, but $\dG$
\beq
\dG(z;\vp)= z\pa_z\udG(z;\vp) = -4cu^2\int_{u^2}^{\infty}dt\;\frac{1}{2^d\pi^{d/2}}\frac{1}{(u^2+t)^{d/2}}e^{-\frac{tu^2}{t+u^2}\vp^2}
\eeq
Defining $t= u^2\tau$, we get
\beq
\dG(z;\vp)=-4cu^{4-d}\int_{1}^{\infty}d\tau\;\frac{1}{2^d\pi^{d/2}}\frac{1}{(1+\tau)^{d/2}}e^{-\frac{\tau}{\tau+1}u^2\vp^2}
\eeq
For $u^2\vp^2 << 1$, the quantity in the exponential is small, because
$$\frac{1}{2}<\frac{\tau}{1+\tau}<1$$
Thus, in the limit $u^2\vp^2 \to 0$, the exponential point-wise (in $\tau$) converges to (and is bounded by) 1. This is also the case for all derivatives of the above function with respect to $u$. Additionally, $\frac{1}{(1+\tau)^{d/2}}$ is integrable on the domain $\tau \in (1,\infty)$. So, using the dominated convergence theorem, we get
\beqn
\dG(z;\vp)&=&-4cu^{4-d}\int_{1}^{\infty}d\tau\;\frac{1}{2^d\pi^{d/2}}\frac{1}{(1+\tau)^{d/2}}\left(1-\frac{\tau}{\tau+1}u^2\vp^2+\frac{1}{2!}\frac{\tau^2}{(\tau+1)^2}u^4\vp^4+\cdots\right)\nonumber\\
&=&-4cu^{4-d}\frac{1}{2^d\pi^{d/2}}\left(I(d;0)-I(d;1)u^2\vp^2+\frac{1}{2!}I(d;2)u^4\vp^4+\cdots\right)
\eeqn
where we have defined
\beq
I(d;m) = \int_1^{\infty}d\tau \frac{\tau^m}{(1+\tau)^{d/2+m}} = \frac{2}{d-2}{} _2F_1\left(\frac{d-2}{2},\frac{d}{2}+m,\frac{d}{2};-1\right)
\eeq
which is well-defined for all $m$ provided $d>2$. The first few of these integrals are given by
\beq
I(d;0) = \frac{2^{2-d/2}}{d-2}
\eeq
\beq
I(d;1) = \frac{2^{1-d/2}(d+2)}{d(d-2)}
\eeq
\beq
I(d;2) = \frac{2^{-d/2}(d^2+6d+16)}{d(d^2-4)}
\eeq
and so on. So, in position space, we get 
\beqn
\dG(z;\vx,\vy) &=& -\frac{4cu^{-2\nu}}{2^d\pi^{d/2}}\left(I(d;0)+I(d;1)u^2\Box_{(x)}+\frac{1}{2!}I(d;2)u^4\Box_{(x)}^2+\cdots\right)\delta^d(x-y)\nonumber\\
&=&-Cz^{-2\nu}\left(1+\alpha z^2\Box_{(\vx)}+\cdots\right)\delta^d(x-y)
\eeqn
where 
$$C= \frac{4cI(d;0)}{2^d\pi^{d/2}}M^{2\nu}, \;\;\; \alpha = \frac{d+2}{2dM^2}>0$$
are constants, and recall that 
$$2\nu = \Delta_+-\Delta_- = (d-4)$$

\subsection{Higher spins}
Now we wish to do the same calculation for generic higher-spin currents. In this case,
\beq
G^{\umu_s,\unu_s}_{(s,s)}(z;\vx,\vy)=\frac{2i}{N}\langle J^{\mu_1\cdots\mu_s}(\vx) J^{\nu_1\cdots\nu_s}(\vy)\rangle_{CFT}
\eeq
Using 
\beq
J^{\mu_1\cdots \mu_s}(\vx) = \phi^*_m(\vx)f^{\mu_1\cdots\mu_s}(\overleftarrow{\pa}_{(x)},\overrightarrow{\pa}_{(x)})\phi^m(\vx)
\eeq
we get in momentum space
\beq
G_{(s,s)}^{\umu_s,\unu_s}(z;\vp) = c_s\int \frac{d^d\vq}{(2\pi)^d}\frac{K(z^2(\vp-\vq)^2/M^2)}{(\vp-\vq)^2}f^{\mu_1\cdots\mu_s}(i\vq,i(\vp-\vq))\frac{K(z^2\vq^2/M^2)}{\vq^2}f^{\nu_1\cdots\nu_s}(-i\vq,-i(\vp-\vq))
\eeq
Once again, using $K(s) = e^{-s}$ and Schwinger parameters, we get
\beq
G_{(s,s)}^{\umu_s,\unu_s}(z;\vp) = c_s\int_{u^2}^{\infty}dt\int_{u^2}^{\infty}ds\int \frac{d^d\vq}{(2\pi)^d}f^{\mu_1\cdots\mu_s}(i\vq,i(\vp-\vq))f^{\nu_1\cdots\nu_s}(-i\vq,-i(\vp-\vq))e^{-s\vq^2-t(\vp-\vq)^2}
\eeq
which can be conveniently written as
\beq G_{(s,s)}^{\umu_s,\unu_s}(z;\vp)=\lim_{\vj\to 0}\int_{u^2}^{\infty}dt\int_{u^2}^{\infty}ds\;f^{\mu_1\cdots\mu_s}\left(\frac{\pa}{\pa{\vj}},i\vp-\frac{\pa}{\pa{\vj}}\right)f^{\nu_1\cdots\nu_s}\left(-\frac{\pa}{\pa{\vj}},-i\vp+\frac{\pa}{\pa{\vj}}\right)\int \frac{d^d\vq}{(2\pi)^d}e^{-s\vq^2-t(\vp-\vq)^2+i\vq\cdot \vj}
\eeq
Upon doing the $\vq$ integration, we get
\beqn
G_{(s,s)}^{\umu_s,\unu_s}(z;\vp)&=&\frac{c_s}{2^d\pi^{d/2}}\;\lim_{\vj\to 0}f^{\mu_1\cdots\mu_s}\left(\frac{\pa}{\pa{\vj}},i\vp-\frac{\pa}{\pa{\vj}}\right)f^{\nu_1\cdots\nu_s}\left(-\frac{\pa}{\pa{\vj}},-i\vp+\frac{\pa}{\pa{\vj}}\right)\nonumber\\ &\times &\int_{u^2}^{\infty}dt\int_{u^2}^{\infty}ds\;\frac{1}{(t+s)^{d/2}}e^{-\frac{ts}{t+s}\vp^2-\frac{1}{4(t+s)}\vj^2+i\frac{t}{t+s}\vp\cdot\vj}
\eeqn
Now taking a $u$ derivative, we see that
\beqn
\dot{G}_{(s,s)}^{\umu_s,\unu_s}(z;\vp)&=&-\frac{2c_su^{4-d-2s}}{2^d\pi^{d/2}}\;\lim_{\vj'\to 0}f^{\mu_1\cdots\mu_s}\left(\frac{\pa}{\pa{\vj'}},iu\vp-\frac{\pa}{\pa{\vj'}}\right)f^{\nu_1\cdots\nu_s}\left(-\frac{\pa}{\pa{\vj'}},-iu\vp+\frac{\pa}{\pa{\vj'}}\right)\nonumber\\ 
&\times& \int_1^{\infty}d\tau\frac{1}{(1+\tau)^{d/2}}e^{-\frac{\tau}{\tau+1}u^2\vp^2-\frac{1}{4(\tau+1)}\vj'^2}\left(e^{\frac{i}{\tau+1}u\vp\cdot\vj'}+e^{\frac{i\tau}{\tau+1}u\vp\cdot\vj'}\right)
\eeqn
where $\vj = u\vj'$. In order to proceed, we need to know the explicit form of $f^{\mu_1\cdots \mu_s}$, and the detailed form of the kernel above will depend on this explicitly. However, in the higher-spin Coulomb gauge we choose the higher-spin fields to be divergenceless, and the only piece of interest is the term proportional to $\eta^{<\mu_1<\nu_1}\cdots \eta^{\mu_s>\nu_s>}$, where the angular brackets refer to the  traceless, symmetric combination. In this case, it is evident for the same reason as in the $s=0$ case, that we have 
\beq
\dot{G}^{\umu_s,\unu_s}(\vx,\vy) = C_sz^{-2\nu}\left(1+\alpha_sz^2\Box_{(\vx)}+\cdots\right)\delta^d(\vx-\vy)\eta^{<\mu_1<\nu_1}\cdots\eta^{\mu_s>\nu_s>}
\eeq
with $\alpha_s>0$.

\providecommand{\href}[2]{#2}\begingroup\raggedright\endgroup

\end{document}